\documentclass[fleqn,usenatbib]{mnras}

\usepackage{newtxtext,newtxmath}
\usepackage[T1]{fontenc}

\usepackage{graphicx}	% Including figure files
\usepackage{amsmath}	% Advanced maths commands
\usepackage{gensymb}
\usepackage{comment}

\usepackage{booktabs}
\usepackage{longtable, lipsum}
\title[Pulsar Polarization]{Pulsar polarization: a broad-band population view with the Parkes Ultra-Wideband receiver}

\author[L. S. Oswald et al.]{L. S. Oswald$^{1,2}$\thanks{E-mail: lucy.oswald@physics.ox.ac.uk (LSO)}, 
S. Johnston$^{3}$, 
A. Karastergiou$^{1}$, 
S. Dai$^{4}$, 
M.~Kerr$^{5}$, 
M.~E. Lower$^{3}$, \newauthor
R.~N. Manchester$^{3}$, 
R. M. Shannon$^{6,7}$, 
C.~Sobey$^{8}$, 
P. Weltevrede$^{9}$
\\
$^{1}$Department of Astrophysics, University of Oxford, Denys Wilkinson Building, Keble Road, Oxford OX1 3RH, UK\\
$^{2}$Magdalen College, University of Oxford, Oxford OX1 4AU, UK\\
$^{3}$Australia Telescope National Facility, CSIRO, Space and Astronomy, PO Box 76, Epping, NSW 1710, Australia\\
$^{4}$School of Science, Western Sydney University, Locked Bag 1797, Penrith, NSW 2751, Australia\\
$^{5}$ Space Science Division, Naval Research Laboratory, Washington, DC 20375, USA\\
$^{6}$ Centre for Astrophysics and Supercomputing, Swinburne University of Technology, PO Box 218, Hawthorn, VIC 3122, Australia\\
$^{7}$ The ARC Centre for Excellence for Gravitational-wave Discovery (OzGrav)\\
$^{8}$ CSIRO Space and Astronomy, PO Box 1130 Bentley, WA 6102, Australia\\
$^{9}$Jodrell Bank Centre for Astrophysics, Department of Physics and Astronomy, University of Manchester, Manchester M13 9PL, UK\\
}

\date{Accepted XXX. Received YYY; in original form ZZZ}

\pubyear{2023}

\begin{document}
\label{firstpage}
\pagerange{\pageref{firstpage}--\pageref{lastpage}}
\maketitle

\begin{abstract}
The radio polarization properties of the pulsar population are only superficially captured by the conventional picture of pulsar radio emission. We study the broadband polarization of 271 young radio pulsars, focusing particularly on circular polarization, using high quality observations made with the Ultra-Wideband Low receiver on Murriyang, the Parkes radio telescope. We seek to encapsulate polarization behaviour on a population scale by defining broad categories for frequency- and phase-dependent polarization evolution, studying the co-occurrences of these categorizations and comparing them with average polarization measurements and spin-down energy ($\dot{E}$). This work shows that deviations of the linear polarization position angle (PA) from the rotating vector model (RVM) are linked to the presence of circular polarization features and to frequency evolution of the polarization. Polarization fraction, circular polarization contribution and profile complexity all evolve with $\dot{E}$ across the population, with the profiles of high-$\dot{E}$ pulsars being simple and highly linearly polarized. The relationship between polarization fraction and circular contribution is also seen to evolve such that highly polarized profiles show less variation in circular contribution with frequency than less strongly polarized profiles. This evolution is seen both across the population and across frequency for individual sources. Understanding pulsar radio polarization requires detailed study of individual sources and collective understanding of population-level trends. For the former, we provide visualizations of their phase- and frequency-resolved polarization parameters. For the latter, we have highlighted the importance of including the impact of circular polarization and of $\dot{E}$.
\end{abstract}


%%%%%%%%%%%%%%%%%%%%%%%%%%%%%%%%%%%%%%%%%%%%%%%%%%

%%%%%%%%%%%%%%%%% BODY OF PAPER %%%%%%%%%%%%%%%%%%

\section{Introduction}

When considering the origins and behaviour of pulsar radio polarization, the results from an observational approach are dependent on the capacities of radio telescope technology. The Ultra-Wideband Low receiver (UWL) on Murriyang, the 64-m Parkes radio telescope, has opened up a new broad-band perspective on pulsar radio emission. With the capacity to observe radio pulsars across a continuous bandwidth from 704--4032~MHz \citep{Hobbs2020}, we can now answer questions on a large scale about the frequency-dependent nature of pulsar radio polarization and what behaviour is seen across the pulsar population. 

The classic, and simplistic, picture of radio pulsar polarization is generally summarized as follows \citep[][]{lorimer2005handbook}: pulsars can be anything from strongly to weakly polarized, that polarization is predominantly linear (average linear polarization fraction of approximately 20\%, but pulsars may be up to 100\% linearly polarized) but most pulse profiles also exhibit a small amount of circular polarization (average absolute circular polarization fraction of approximately 10\%). The observed angle of the linear polarization (position angle or PA) evolves smoothly across the pulse profile following an S-shaped curve, which can be described as a geometrical effect of the magnetic field according to the rotating vector model \citep[RVM,][]{Radhakrishnan1969}. Fits of the RVM to PA profiles are often used to determine pulsar geometries---the inclination of the magnetic field from the spin axis and the impact parameter of the line of sight \citep[see for example][]{Johnston2019a}. 

A further dimension to observable pulsar polarization is that of observing frequency. According to the RVM, the PA profile results purely from the geometry of the pulsar and the observer's viewing angle, factors which are independent of frequency. However, it has long been known that this does not fully account for the observational picture of pulsar polarization, since individual pulsars show a wide variety of other effects that need to be considered. For example, pulsars are generally seen to depolarize at high frequencies \citep[e.g.][]{Sobey2021a} and some pulsars show a frequency-dependent transition from linear to circular polarization \citep{VonHoensbroech1999b}. The observation that many pulsar intensity profile shapes are observed to evolve with frequency could be explained as originating from refractive effects in the magnetosphere \citep{Lyubarskii1998a}. The idea of radius-to-frequency mapping \citep[RFM,][]{Ruderman1975}---that there is a relationship between radio frequency and the height of emission above the pulsar surface---is also commonly invoked as an explanation.

An additional level of complexity to consider is the presence of two orthogonally polarized modes of emission in pulse profiles. These are most clearly identifiable by the presence of orthogonal jumps in the PA of 90$\degree$, at a phase bin at which the total polarization drops to zero \citep[e.g.][]{Manchester1975}. These orthogonal jumps are identified as being the pulse phase at which mode dominance is exchanged, such that the stronger mode and the weaker mode contributing to the observed profile swap. 
\cite{Karastergiou2005} studied the frequency-dependence of linear polarization for 48 pulsars and demonstrated that the observed frequency evolution could be explained by the pulse profiles being composed of orthogonally polarized modes that had different spectral indices for their intensities. \cite{Smits2006} showed that pulse profile components dominated by a single orthogonal mode tended to increase in intensity relative to the rest of the profile with increasing frequency. Regardless of the precise mechanisms behind emission heights and propagation paths of pulsar radio emission, interactions between orthogonal modes are by definition path-dependent and may therefore be the cause of phase- and frequency-dependent evolution of polarization fractions and non-RVM features in the PA.

The orthogonally polarized modes of emission are explained as originating from birefringence in the magnetospheric plasma. It is theorised that the ordinary (O) and extraordinary (X) modes propagate along different paths through the magnetosphere, and the observed polarization is set at the polarization limiting radius, the point at which the radiation is decoupled from the magnetosphere \citep[e.g.][]{Barnard1986}. This theory goes some way towards explaining the diversity of polarization behaviour of pulse profiles across the pulsar population. But there remain two key aspects of pulsar polarization that need to be addressed: namely that most pulsars exhibit at least a small amount of circular polarization, and that many pulsars display polarization properties, such as features in the PA profiles, that are not adequately explained by the RVM plus orthogonal jumps. The wide variety of features seen in the population---and the difficulty of studying more than a few pulsars in detail at a given time---means that the more in-depth studies of radio pulsar polarization have generally been limited to a small number of pulsars, or even a single pulsar, as for the example of PSR~B1451$-$68 \citep{Dyks2021a}. Another recent case of interest is that of PSR~B1919$+$21, for which \cite{Primak2022} studied the polarization of its drifting sub-pulses. They found that some phase regions of the pulse profile appeared to result from incoherently superposed orthogonally polarized modes, and other regions resulted from superposition that is partially coherent. Further, the coherently superposed regions exhibited rotation of the polarization vectors about the Poincar\'e sphere at the same rate as the drifting subpulse intensity modulation. 

A coherent or partially-coherent model of orthogonal mode superposition explains polarization behaviour well in these individual cases, but the population-wide ramifications of such a model have not previously been addressed observationally. Nevertheless, there is growing evidence in large scale pulsar studies of the need to address the magnetospheric origins of more complex polarization behaviour \citep[e.g.][]{Ilie2019}. Broad-band radio observations, made with new and upgraded instruments, mean that it is now possible to measure the continuous evolution of polarization across a large fractional bandwidth for a large number of pulsars: see for example the observations below 200~MHz made with LOFAR \citep{Noutsos2015}. This large increase in the parameter space of polarization measurements means we can clearly identify how polarization behaviour varies across both the pulsar population and across frequency for each pulsar, giving new insight into the question of how it is generated.

In this paper we present full polarization observations of a large sample of non-recycled pulsars observed with the UWL receiver on Murriyang, the Parkes radio telescope, with the goal of updating our understanding of the behaviour of pulsar polarization in broad-band observations. \cite{Johnston2021} introduced the initial results of two years of such observations and \cite{Sobey2021a} presented a polarization census for 40 bright pulsars from this dataset. The work presented here is that previewed and promised by \cite{Oswald2020}, where we presented plans for extending our broad-band polarization study to a population-level scale, with particular focus on the frequency dependence of circular polarization and the relationships between polarization behaviour and pulsar parameters including spin-down energy $\dot{E}$.

We describe our observations and data analysis methods in Sections \ref{sec:obs} and \ref{sec:data_analysis_methods} respectively. In Section \ref{sec:data_results}, we present visualizations of how each pulsar's phase-resolved polarization properties evolve with frequency. 
We also categorize the key qualitative polarization behaviours of the pulsar population, and investigate co-occurrences of these features and their evolution with frequency across the population, with particular focus on circular polarization and on $\dot{E}$. A summary of the results is presented in Section \ref{sec:summary} and the full collection of broad-band visualizations of the dataset are presented as supplementary material online.

\section{Observations}
\label{sec:obs}

Polarization data were obtained under the P574 observing programme, ongoing since 2007 at the Parkes radio telescope, Murriyang. Prior to 2018 November, observations were predominantly carried out monthly at 1369~MHz with a bandwidth limited to 256~MHz with occasional observations at 3100~MHz and 700~MHz \citep[see][]{Johnston2018, Petroff2013}. For this paper we use observations from 2018 November onwards which used the UWL receiver \citep{Hobbs2020}. The observational band runs from 704~MHz to 4032~MHz, with 3328 channels each of width 1~MHz. The data are coherently dedispersed in each channel, folded at the topocentric pulsar period and collated into sub-integrations of length 30~s with 1024 phase bins per period for the duration of the observation (typically 200~s). Full Stokes information is recorded and the data is recorded to disk in {\sc psrfits} format \citep{Hotan2004, VanStraten2010}.

Data processing is carried out within {\sc psrchive} \citep{Hotan2004}. A key element of the analysis is the removal of radio-frequency interference (RFI) from the data which is carried out using a median filtering process in both frequency and time. Typically, approximately 1200 channels (out of 3328) are discarded from the observations. Polarization and flux calibration is carried out via the routine {\sc pac} using observations of a pulsed calibration signal and observations of PKS~B1934--68. Corrections due to impurities in the receiver are computed via observations of the pulsar PSR~J0437--4715 and also applied \citep{vanStraten2013}. Further details of the observing programme with the UWL can be found in \cite{Johnston2021}.

Individual observations are short so in order to improve the signal-to-noise ratio (S/N) in the pulsar profile we summed together all observations taken over a three year interval (approximately 35 epochs per pulsar) so that the typical total observation time is some 7000~s. The frequency resolution of the UWL is 3328 channels across the band. To increase S/N, and for consistency across the sample, we split the whole band into eight sub-bands and generated a single pulse profile per sub-band. For this study we use a dataset of profiles from 271 pulsars observed regularly as part of the observing programme.

\begin{figure}
    \centering
    \includegraphics[width=\columnwidth]{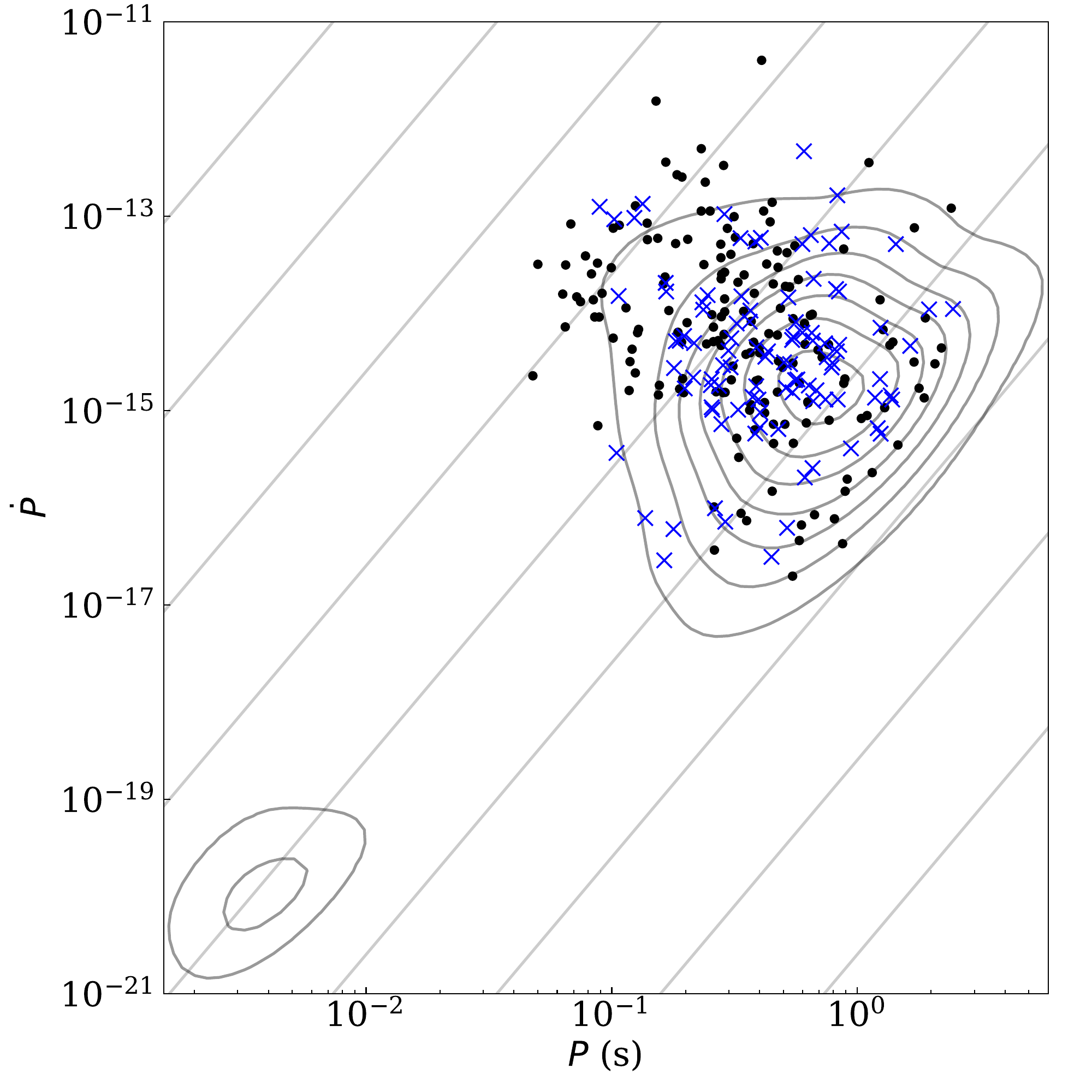}
    \caption{Positions on the $P$--$\dot{P}$ diagram of the 271 pulsars presented in this dataset. Blue crosses: 101 high-S/N, non-scattered pulsars used in categorization analysis. Black points: remaining 170 pulsars of the dataset. Contours: Gaussian kernel density estimation of the pulsar population as taken from {\sc psrcat}. Also plotted are diagonal lines of constant $\dot{E}$, from $10^{26}$~erg~s$^{-1}$ (bottom right) to $10^{42}$~erg~s$^{-1}$ (top left) in increments of $10^{2}$~erg~s$^{-1}$. }
    \label{fig:PPdot}
\end{figure}

Fig. \ref{fig:PPdot} shows the positions of our pulsar sample on the $P$--$\dot{P}$ diagram in relation to the population distribution given by the ATNF catalogue {\sc psrcat} \citep{Manchester2005}, version 1.68\footnote{\href{https://www.atnf.csiro.au/research/pulsar/psrcat}{https://www.atnf.csiro.au/research/pulsar/psrcat}}, accessed via queries with {\sc psrqpy} \citep{Pitkin2018}. The pulsars in our sample are marked as points and crosses and the ATNF population is indicated by contour lines obtained by fitting a Gaussian kernel density estimate to the distribution of pulsars. As can be seen, our sample covers most of the area of $P$--$\dot{P}$ space occupied by the non-recycled population, but with a distribution that favours higher $\dot{E}$ values than that of the catalogue population and no pulsars with $\dot{E}\leq 10^{30}$erg~s$^{-1}$.

\section{Data analysis methods}
\label{sec:data_analysis_methods}

\subsection{Visualization}

\begin{figure*}
    \centering
    \includegraphics[width=\textwidth]{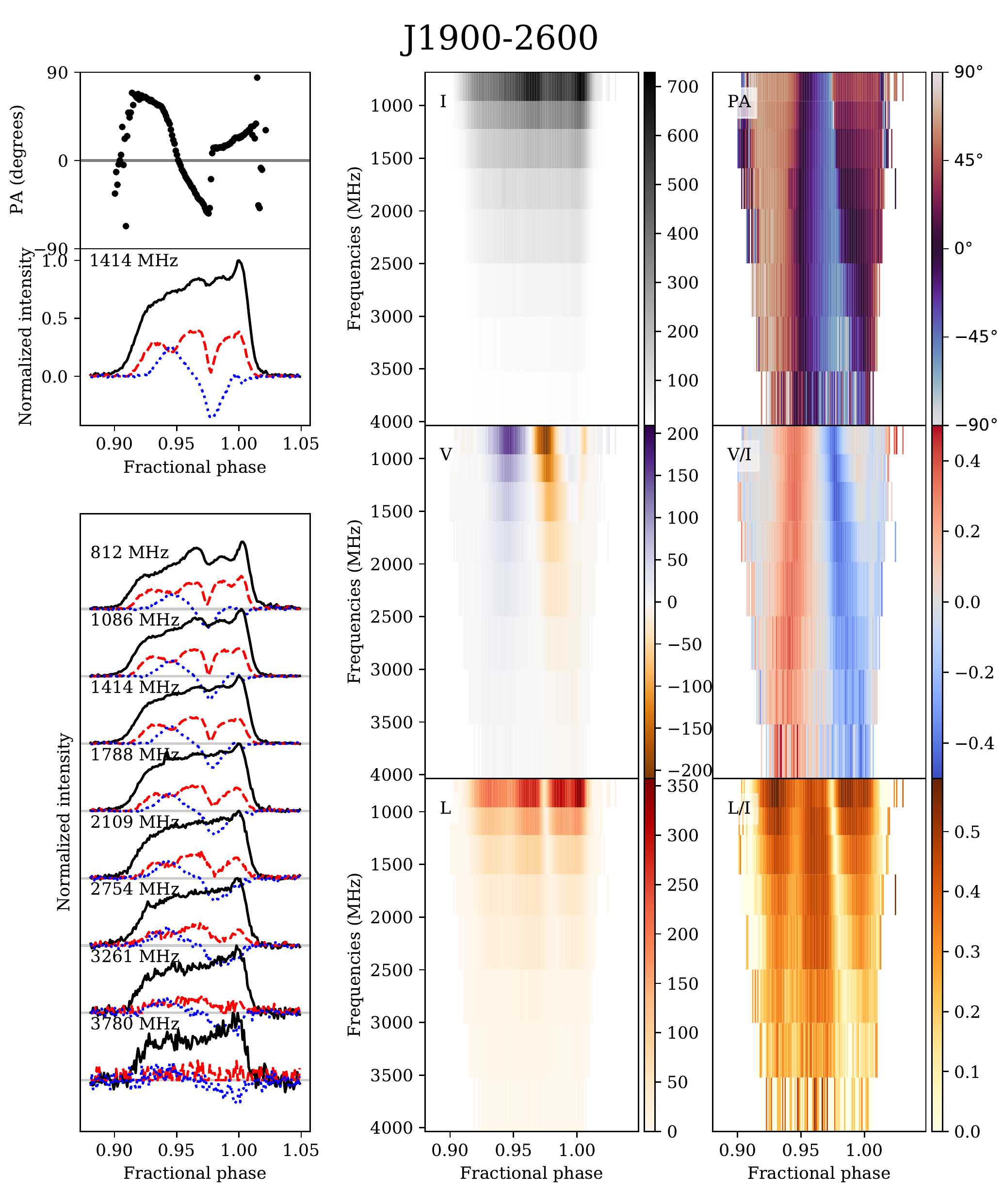}
    \caption{Broad-band visualization of PSR~J1900$-$2600 observed in fold with the UWL on Murriyang, the Parkes Radio Telescope. Top left: Pulse profile at 1414~MHz, with total intensity (black), linear polarization (red dashes), circular polarization (blue dots) and position angle (PA, black points). Left: Pulse profiles at eight frequencies from 812--3780~MHz, colour scheme as before. Right: Waterfall plots of phase- and frequency-resolved polarization showing, from top left to bottom right, total intensity, PA, circular polarization, fractional circular polarization, linear polarization, fractional linear polarization. The PA colour scheme is chosen to be cyclical around 180$\degree$ to account for PA wrapping. The circular polarization colour scheme is chosen so that purple is positive and orange is negative, and the fractional circular polarization colour scheme is chosen so that red is positive and blue is negative. Colour scale units for $I$, $V$ and $L$ are in mJy. The waterfall plot polarizations are only shown where the S/N of the total intensity exceeds 5; the same is true for the PA profile in the top left subplot. }
    \label{fig:ExamplePulsar}
\end{figure*}

We present the results of our observations and data processing as frequency- and phase-resolved waterfall plots for each of our 271 pulsars. These are presented in the online supplementary material (file name: ``Appendix B Polarization visualizations for pulsar sample'', pulsars presented in numerical order from PSR~J0034$-$0721 to J2346$-$0609). An example for PSR~J1900$-$2600 is shown in Fig. \ref{fig:ExamplePulsar}. For the 15 pulsars that have an interpulse---a second pulse at half the pulse period, inferred to originate from the other magnetic pole of the neutron star---we present the same information for the interpulse as for the main pulse. We selected the waterfall plot presentation to display the maximum possible information about the key observables of each pulsar, and in particular, the smooth frequency evolution of many features in the pulsar polarization. 

In each visualization, we show a standard pulse profile plot, with linear and circular polarization and PA profile at 1400~MHz, in the top left corner. Below it, we show a series of these pulse profiles across the eight sub-bands. Finally, we display a series of six waterfall plots showing, from top left to bottom right, total intensity $I$, PA, circular polarization $V$, fractional polarization $V/I$, linear polarization $L$ and fractional linear polarization $L/I$. For all of these visualizations the linear polarization $L$ has been bias-corrected according to the prescription of \cite{Everett2002}. The purpose of having both polarized flux and fractional polarization plots for linear/circular polarization is that, whereas the polarized flux plots show the evolution of polarization structures with frequency and phase, the fractional polarizations show the extent to which these structures are tied to total intensity.

For the waterfall plots, we only show the pulse profile where the S/N of Stokes $I$ is greater than 5. This is also the case for the PA in the standard pulse profile plot in the top left. This means that PA values are still being calculated at phases with strong intensity but little to no linear polarization. However, the goal of data visualization is different from that of data analysis. We choose to maintain the same S/N cut-off for all variables: partly for consistency and partly because we find that this provides the maximum visual information about frequency-dependent evolution even at very low linear polarization.

In order to define the S/N of the intensity profile, $I/\sigma$, we use an iterative approach to remove outlier points (on-pulse region and any residual impulsive RFI) to converge onto a standard deviation measurement of the off-pulse region. We use this method to calculate $\sigma$ for all pulsars other than PSRs~J0820$-$4114 and J1834$-$0426. These pulsars have very wide profiles, causing the iterative outlier-removal method to fail. In those cases we define the off-pulse region by eye. We split the off-pulse region into four sections, calculate the standard deviation of each section and select the smallest as the least likely to include the effect of impulsive RFI and therefore the most representative value of $\sigma$. 

We also present a second piece of online supplementary material visualizing two additional polarization parameters. The first of these is the total polarization $P$ as a fraction of the total intensity, Stokes $I$: $p = P/I$. The second is the ``circular contribution'', $\theta$. We introduce the concept of circular contribution as a helpful metric for investigating the relative levels of linear and circular polarization in a pulse profile. We define $\theta$ relative to the ellipticity angle $\chi$ as $\theta = 2\lvert\chi\rvert$. It is the magnitude of the angle of latitude on the Poincar\'e sphere, given by the equation $\theta = \arctan{(\lvert V\rvert/L)}$. It is helpful to define the parameter in this way, rather than using ellipticity directly, because it gives a more clearly comparable link between absolute levels of linear and circular polarization, and their visualization on the Poincar\'e sphere. Fully linearly polarized radiation will have a circular contribution of $\theta = 0\degree$, whilst fully circularly polarized radiation has the maximum circular contribution of $\theta = 90\degree$. 

This second set of figures in the online supplementary material (file name: ``Appendix C Visualizations of p and theta for pulsar sample'') show waterfall plots of $P/I$ and $\theta$, following the same S/N and bias correction steps as described for the waterfall plots in Fig. \ref{fig:ExamplePulsar} and in the first piece of supplementary material (``Appendix B Polarization visualizations for pulsar sample'').

\subsection{Polarization measurements}
\label{sec:polmeas}

In order to make inferences about the polarization behaviour of the pulsar population we need to reduce the dimensionality of this observational information. We take two approaches to address this: polarization measurements and categorization of polarization behaviour. For the first of these, we sum the Stokes and polarization parameters across phase to obtain a single characteristic value per pulse profile, per frequency. To avoid off-pulse RFI biasing these values, we sum only the bins for which the intensity S/N $>3$ that lie within our visually-defined on-pulse region.

We calculate two sets of average polarization fractions for each pulsar. For the first set, we calculate total, linear and absolute circular polarizations from the Stokes parameters for each pulse profile and then sum these across phase. We label polarization fractions with lower case letters $p$, $l$ and $\lvert v\rvert$, and for this first set, we indicate that these are averages by marking them with bars: $\overline{p}$ etc. For the second set, we sum Stokes parameters across phase first and then convert these into polarization fractions. We label this second set of average polarization fractions with asterisks. This gives the following formulae:
\begin{align*}
    \overline{p} &= \frac{\Sigma \left(\sqrt{Q^{2} + U^{2} + V^{2}}^{BC}\right)}{\Sigma I} \\ 
    \overline{l} &= \frac{\Sigma \left(\sqrt{Q^{2} + U^{2}}^{BC}\right)}{\Sigma I} \\ 
    \overline{\lvert v\rvert} &= \frac{\Sigma \left(\lvert V\rvert^{BC}\right)}{\Sigma I} \\
    p^{\ast} &= \frac{\sqrt{(\Sigma Q)^{2} + (\Sigma U)^{2} + (\Sigma V)^{2}}}{\Sigma I} \\
    l^{\ast} &= \frac{\sqrt{(\Sigma Q)^{2} + (\Sigma U)^{2}}}{\Sigma I} \\
    \lvert v\rvert^{\ast} &= \left\lvert\frac{\Sigma V}{\Sigma I}\right\rvert
\end{align*}
where $\Sigma$ indicates summing across phase and the superscript $^{BC}$ indicates that bias correction is applied to the polarizations before summing. We also use the same bar/asterisk labelling for $\theta$ when describing variables summed over pulse phase.  Uncertainties are inferred using the previously described measurements of $\sigma$ and standard error propagation. We bias-correct the linear polarization as described in \cite{Everett2002} and use the same method for the total polarization as well. To bias correct the absolute circular polarization $\lvert V\rvert$ we follow the method of \cite{Karastergiou2003a} and \cite{Posselt2022y} and subtract the expectation value of the distribution. Being a half-normal distribution, this is done as follows: 
\begin{equation}
\lvert V\rvert^{BC}=\left\{ 
\begin{array}{ll}
\lvert V\rvert - \sigma \sqrt{\frac{2}{\uppi}} & \mbox{if $\lvert V\rvert > \sigma \sqrt{\frac{2}{\uppi}}$} \\ 
0 & \mbox{otherwise,}\end{array} \right.
\end{equation}
where $\sigma$ is the standard deviation of the intensity off-pulse region as described above. 

The polarization measurements $\overline{l}$, $\overline{\lvert v\rvert}$, $\overline{p}$ and $\overline{\theta}$, along with their uncertainties, are listed for each pulsar in Table \ref{tab:all_ave_psr_results}. For pulsars with interpulses, we calculate the interpulse polarization fractions in the same way and list them separately in Table \ref{tab:all_ave_psr_results}.

\subsection{Categorization of profile polarization features}
\label{subsec:polcategories}

We seek to present a summary of key features and trends related to polarization structure observable in the pulsar population. In addition to the average numerical values, we also categorize certain polarization and profile features of a high-S/N subset of 101 pulsars. We apply a S/N cut to identify which pulsars to take forward for this further analysis: we keep those pulsars for which the peak intensity S/N $> 10$ for all eight frequency sub-bands. We define the peak intensity S/N by taking the maximum intensity value in the on-pulse region, subtracting the median of the off-pulse region to correct any zero-offset of the baseline, and then dividing by $\sigma$. We also identify any strongly scattered pulsars by eye and remove them from the high-S/N sample (all scattered pulsars are marked with an asterisk in Table \ref{tab:all_ave_psr_results}). The reasoning for this is that if scattering is strong enough to be clearly visible by eye, then the effect of scattering is likely to impact the features of the polarization profile observed \citep{Karastergiou2009a}.

We identify polarization features by eye from the broad-band waterfall plots generated for this dataset. The objective is not to state absolute numerical results for these categorizations, many of which are by definition difficult to define unequivocally for every pulsar, and in any case such precise quantification would not provide much additional information based on what we wish to describe. Instead, we seek to obtain a general understanding of the existences of these features and the ways in which they co-occur. For simplicity, we define all but one of the categories as Boolean yes-no cases: either a pulsar exhibits a particular feature or it does not. The categories are defined as follows. 

\begin{enumerate}
\item Visible frequency evolution of... 
\begin{enumerate}
    \item ...intensity profile shape
    \item ...PA profile shape
    \item ...linear polarization fraction
    \item ...linear polarization profile shape
    \item ...circular polarization fraction
    \item ...circular polarization profile shape
\end{enumerate}
\item Phase variation of polarization: 
\begin{enumerate}
    \item ``True'' orthogonal jump (Orthogonal jump in PA at the phase at which total polarization drops to zero)
    \item Orthogonal jump in PA with non-zero circular polarization
    \item PA deviation from RVM not otherwise accounted for by the two previous categories
    \item $V$ changes hand with phase
\end{enumerate}
\end{enumerate}

Our only non-Boolean category is \textit{Profile Complexity}. Integrated pulse profiles are observed to have a wide range of shapes, with multiple components that may be blended together. We choose to investigate the scaling of polarization complexity by assigning each pulsar to one of three categories. We number these in order of increasing complexity: category 0 describes profiles with a single component. Profiles with a small number of components (two to four) that are well separated and/or symmetric, or a strongly asymmetric single component, are assigned to category 1. Category 2 comprises complex profiles with multiple components or blended components. The categorization of any individual pulsar is on some level subjective, but we find that the three-category set-up enables clearer distinctions to be drawn between the pulsars in categories 0 and 2.  

It is worth making clear the difference between categories (i)c/e and (i)d/f, that is to say the difference between polarization fraction and polarization profile shape. For the former, we looked for a change across frequency of the fraction of the linear/circular polarization relative to total intensity in at least part of the pulse profile. For the latter, we looked not at the polarization fraction, but at the shapes of any distinct features in the polarized part of the pulse profile and whether these evolved with frequency. PSR~J1900$-$2600, shown in Fig. \ref{fig:ExamplePulsar}, has frequency evolution of both the polarization fraction and profile shape of its linear polarization. The former can be seen in the stacked pulse profiles: the linear polarization fraction decreases significantly at high frequencies. The latter is best identified in the $L/I$ plot, where the shifting positions of light and dark regions of the profile indicate a change in shape of the linear polarization profile across frequency.

It is important that the frequency evolution of polarization profile shape is defined with respect to the fractional polarization plot (e.g. $L/I$) rather than the absolute polarization (e.g. $L$) because this separates polarization evolution as distinct from evolution of the total intensity profile. 
A relevant example to consider is PSR~J1048$-$5832, the visualizations for which can be found in the online supplementary material. This is a pulsar for which the linear polarization profile is distinctly different from the intensity profile, as can be seen from the waterfall plots of $I$ and $L$, yet the polarization fraction $L/I$ does not evolve with frequency. We also note that, since pulsars tend to have weaker circular polarization than linear polarization, this may impact the relative count of visible features in circular polarization, however, since we apply this analysis only to high-S/N pulsars, we expect this effect to be small.

The reason for separating PA non-RVM behaviour into three categories is as follows. Following the theoretical expectation that an orthogonal jump takes place when there is a change in dominance between the two orthogonal modes of emission, we would expect ``true'' orthogonal jumps to be accompanied by a drop of the polarization to zero. As described by \cite{Dyks2020}, apparent orthogonal jumps in PA will also result from the polarization state transitioning across the Poincar\'e sphere and passing close to the Stokes $V$ pole. In these cases, the polarization does not drop to zero, instead it is rotated from $L$ into $V$. This means that there will be significant measured circular polarization accompanying the PA jump. Cases where the PA deviates from an expected RVM track but do not show a clear jump are the hardest to define: is this behaviour similar to either of the jump behaviours or is it fundamentally different? We seek to categorize the different observational manifestations so that we can compare their co-occurrences with other categories. %
and so determine whether the apparent separation between categories is indeed meaningful. 

The full categorization for each pulsar is given in Table \ref{tab:all_ave_psr_results}, where each category is labelled as in this list. Note that for pulsars with interpulses we only categorize the main pulse. 
As an example, PSR~J1900$-$2600 in Fig. \ref{fig:ExamplePulsar} is classified as having the following polarization features: (i)abcdf;(ii)bcd;2. This means it displays frequency evolution of intensity profile, PA profile, linear polarization fraction, linear polarization profile shape and circular polarization profile shape - `(i)abcdf'. It also exhibits a PA jump with non-zero circular polarization at the jump phase, deviation from a classic RVM swing and a change in hand of the circular polarization - `(ii)bcd'. Finally, it has been categorized as having Profile Complexity `2' - multiple components or blended components.

\section{Polarization properties of the dataset}
\label{sec:data_results}
 
Fig. \ref{fig:ExamplePulsar} shows PSR~J1900$-$2600 as an example of how we visualize the polarization properties of each pulsar in the dataset, as described in Section \ref{sec:data_analysis_methods}. Unlike the majority of past publications of pulse profiles, which show only the standard profile with PA, this visualization follows the techniques set out in \cite{Oswald2020} to display the phase- and frequency-evolution of pulse profiles simultaneously, in order to capture the full capability of the UWL. Here, we extend those techniques beyond total intensity and PA profile to encompass polarization fractions, and particularly circular polarization. 

The full collection of data visualizations of 271 pulsars is presented in the online supplementary material. In the remainder of this section we focus on the results of our measurements and categorizations taken collectively. We aim to reduce the number of degrees of freedom required to encapsulate the dataset so that clear and simple statements may be made about its polarization behaviour.

\subsection{Average polarization values}
\label{sec:avepols}

Using the standard measurements for fractional polarizations at 1400~MHz, we find that, on average (median), the pulsars in our sample are 28\% linearly polarized, 5\% circularly polarized and 32\% polarized in total (see Table \ref{tab:polfracs}). This is comparable to general expectations from the literature: \cite{Gould1998} quote 20\% linear and 10\% absolute circular polarization fractions. 

However, careful consideration should be made about the meaning of these values. 

A median does not capture range of polarization fractions observed, or the relationship between polarization fraction and $\dot{E}$ \citep[e.g.][]{Weltevrede2008}, which we discuss further in Section \ref{sec:count_results}. We should also consider the choice of measurement technique. The numbers quoted above are summed polarizations divided by summed intensity ($\overline{p}$, $\overline{l}$ and $\overline{\lvert v\rvert}$). Summing Stokes parameters before converting to polarization ($p^{\ast}$ etc.) results in much smaller polarization fractions. This is expected, since the PA varies across the pulse profile, so that summing Stokes $Q$ and $U$ will have an averaging effect. Similarly, summing Stokes $V$ across phase will generate a smaller measurement if the direction of circular polarization changes hand. 

It is usual in the literature to quote the barred values ($\overline{p}$ etc.), which are a more accurate representation of pulsar polarization fraction. However, it is worth stressing the difference in results for these two sets of parameters for two reasons. First, the starred polarization fractions are the more relevant when searching for pulsars as polarized sources in the image plane \citep[e.g.][]{Sobey2022}. Secondly, the difference between $\overline{p}$ and $p^{\ast}$ is an indicator of phase evolution of polarization in a pulse profile. We note that there is a bigger difference between $\overline{p}$ and $p^{\ast}$ for more complex pulse profiles (category 2 vs. category 0), implying that the presence of additional profile components leads to increased polarization complexity, and/or that pulsars with simpler pulse profiles tend to have flatter PA profiles.

\begin{table}
\caption{Distributions of the measured polarization fractions and circular contribution at 1400~MHz, summarized with the 25th, 50th and 75th quantiles read from left to right. The parameters $l$, $\lvert v\rvert$, $p$ and $\theta$ are as described in Section \ref{sec:data_analysis_methods}, and the measurements are shown for phase-averaged polarizations (barred) and polarizations calculated with phase-averaged Stokes parameters (starred). The 50th quantile is indicated in bold.}
\label{tab:polfracs}
\centering
\begin{tabular}{lll}
\toprule
{} &            Barred &           Starred \\
\midrule
$l$     &  0.17, \textbf{0.28}, 0.52 &  0.10, \textbf{0.19}, 0.43 \\
$\lvert v\rvert$     &  0.03, \textbf{0.05}, 0.09 &  0.02, \textbf{0.05}, 0.10 \\
$p$     &  0.21, \textbf{0.32}, 0.56 &  0.11, \textbf{0.21}, 0.46 \\
$\theta$ ($\degree$) &   6.4, \textbf{11.6}, 18.9 &   6.9, \textbf{13.4}, 28.5 \\
\bottomrule
\end{tabular}
\end{table}

\subsection{Categorization results}
\label{sec:cat_results}

\begin{table}
\caption{Table of percentage counts of how many pulsars fit into each of the categories defined in Section \ref{subsec:polcategories}. The categorization was applied to the 101 pulsars that exceeded the S/N cut defined in Section \ref{sec:data_analysis_methods} and were identified as not being visibly scattered. A horizontal line separates the frequency-dependent and phase-dependent categorizations. }
\label{tab:percentages}
\centering
\begin{tabular}{ll}
\toprule
Polarization category            & Occurrence \\ \midrule
(i)a: $I$ shape frequency evolution    & 74\%       \\
(i)b: PA frequency evolution           & 51\%       \\
(i)c: $L$ fraction frequency evolution & 40\%       \\
(i)d: $L$ shape frequency evolution    & 31\%       \\
(i)e: $V$ fraction frequency evolution & 32\%       \\
(i)f: $V$ shape frequency evolution    & 32\%       \\
\midrule
(ii)a: PA orthogonal jump               & 37\%       \\
(ii)b: PA jump, $\lvert V\rvert > 0$    & 20\%       \\
(ii)c: PA RVM departure                 & 45\%       \\
(ii)d: $V$ change of hand               & 71\%       \\ 
\bottomrule
\end{tabular}
\end{table}

Table  \ref{tab:percentages} shows percentages of the high-S/N sample of 101 pulsars that exhibit each of the characteristics defined in Section \ref{subsec:polcategories}. Most pulsars show clearly visible frequency evolution of profile shape in Stokes $I$ and about half of the sample show frequency evolution of the PA. Frequency evolution of linear and circular polarization is also commonly observed. By far the most common phase-dependent feature identified was there being at least one change of handedness in circular polarization. A sizable fraction of pulsars show orthogonal jumps in their PA profiles, but at least 45\% of pulsars show some sort of departure from RVM behaviour in their PA profiles across phase that cannot be explained as being due to orthogonal jumps. 

\begin{figure*}
    \centering
    \includegraphics[width=\textwidth]{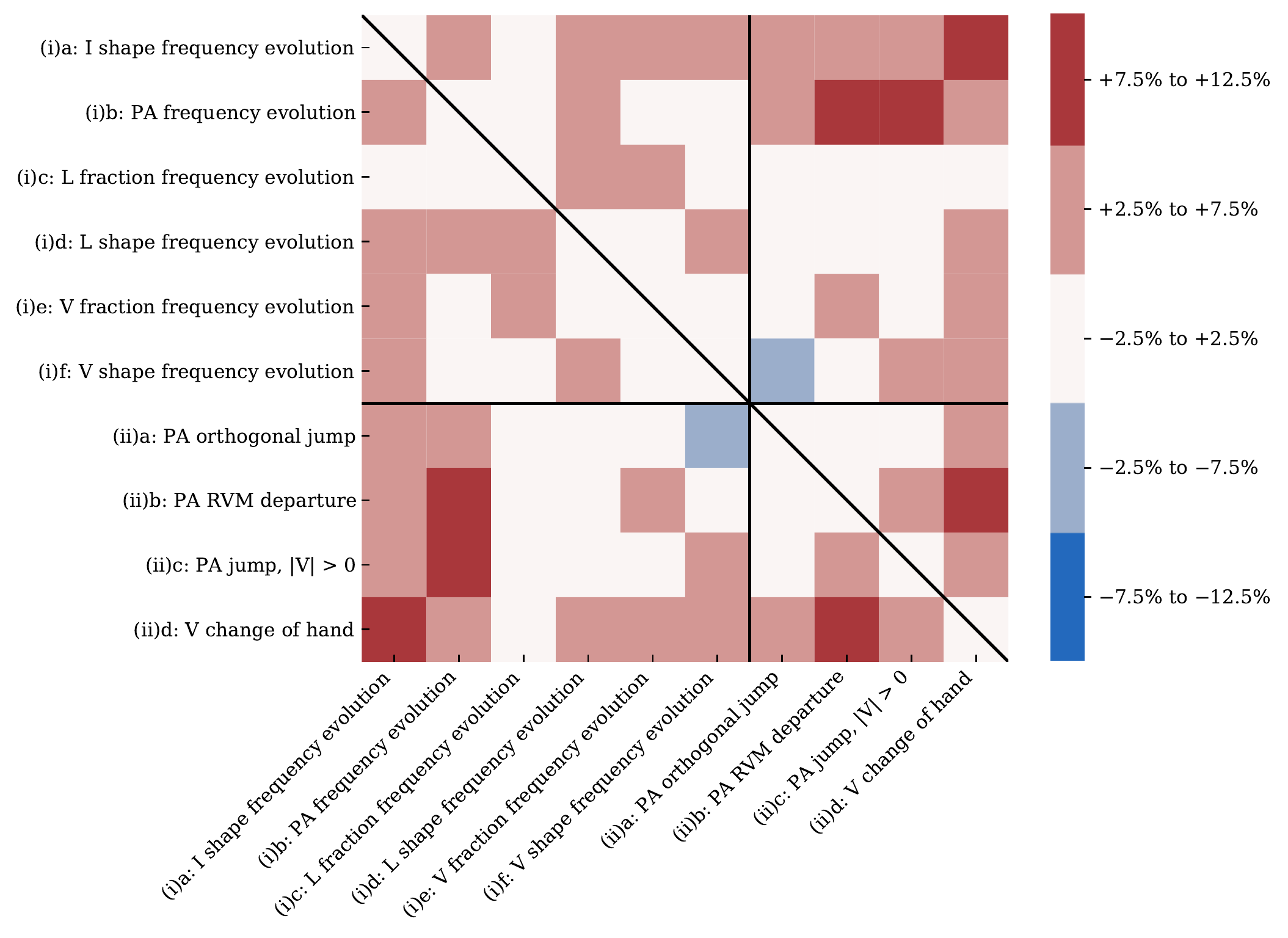}
    \caption{Heat map of the co-occurrence excesses of the categories assigned to the 101 high-S/N pulsar subset, as described in the text and introduced in Table~\ref{tab:percentages}. These are defined as the percentage excesses of categories appearing together in the sample, relative to the expected co-occurrence for independent categories, rounded to the nearest 5\%. The diagonal line indicates the symmetry of the heatmap. Further details are given in the text.}
    \label{fig:heatmap}
\end{figure*}

Having identified these categorizations, the next question to consider is how the occurrences of these profile features relate to each other. For example, we might expect that the pulsars exhibiting non-RVM behaviour are also likely to show frequency-evolution of the PA. From Table \ref{tab:percentages}, we see that roughly half of pulsars show frequency evolution of the PA, and just under half show departures from the RVM. If these two behaviours are independent, then we would expect around a quarter of the sample to exhibit both behaviours. Instead, we find that the two categories co-occur in over a third of the pulsars: an excess of around 10\%. We calculate these co-occurrence excesses for every pair of parameters, round the values to the nearest 5\% and plot them in a heatmap in Fig. \ref{fig:heatmap}.

\subsubsection{Positive and negative associations}
The heatmap in Fig. \ref{fig:heatmap} is divided into sections with black lines indicating the frequency-dependent and phase-dependent categories. The top right and bottom left sections both show a large number of co-occurrences, indicating that there is a link between phase-dependent and frequency-dependent polarization behaviour.

The strongest positive associations between pairs of parameters, marked dark red on Fig. \ref{fig:heatmap}, are: 
\begin{itemize}
    \item PA frequency evolution with PA RVM departures
    \item PA frequency evolution with PA jumps where $\lvert V\rvert > 0$
    \item $V$ change of hand with frequency evolution of total intensity
    \item $V$ change of hand with PA RVM departures. 
\end{itemize}
The only negative association, implying that these two characteristics are found together less often than expected for independent variables, is that between PA orthogonal jumps and $V$ shape frequency evolution. Conversely, $V$ shape frequency evolution does show a link with PA jumps where the circular polarization is non-zero.

\subsubsection{Most and fewest associations}
The two characteristics with the most associations are $I$ shape frequency evolution and $V$ change of hand. These characteristics display some association with all of the categories other than $L$ fraction frequency evolution, which itself is the characteristic with the fewest associations. It only shows a notable link to $L$ shape frequency evolution and to $V$ fraction frequency evolution.

\subsubsection{PA behaviour and circular polarization}
Non-RVM behaviour of the PA is associated more with circular polarization than with linear polarization: PA RVM departures are associated with $V$ fraction frequency evolution and $V$ change of hand, but not with any frequency evolution of linear polarization. Similarly, PA jumps with non-zero circular polarization are associated with $V$ shape frequency evolution and also with $V$ change of hand. Both these associations are weak, but present, whereas there is no association with linear polarization frequency evolution. Taken collectively, the various associations with categories related to circular polarization indicate a strong link between polarization behaviour not explained by the rotating vector model and the presence of circular polarization in the pulse profile. A similar link is noted by \cite{Johnston2022y}, who find that, when fitting the RVM to the PA, a large fraction of the pulsars for which this fails have a circular polarization fraction greater than the linear polarization fraction. This can be explained as resulting from the coherent addition of orthogonal modes, a concept we will discuss further in Section \ref{sec:summary}.

\subsection{Average polarizations, profile complexity and $\dot{E}$}
\label{sec:count_results}

\begin{figure*}
    \centering
    \includegraphics[width=0.8\textwidth]{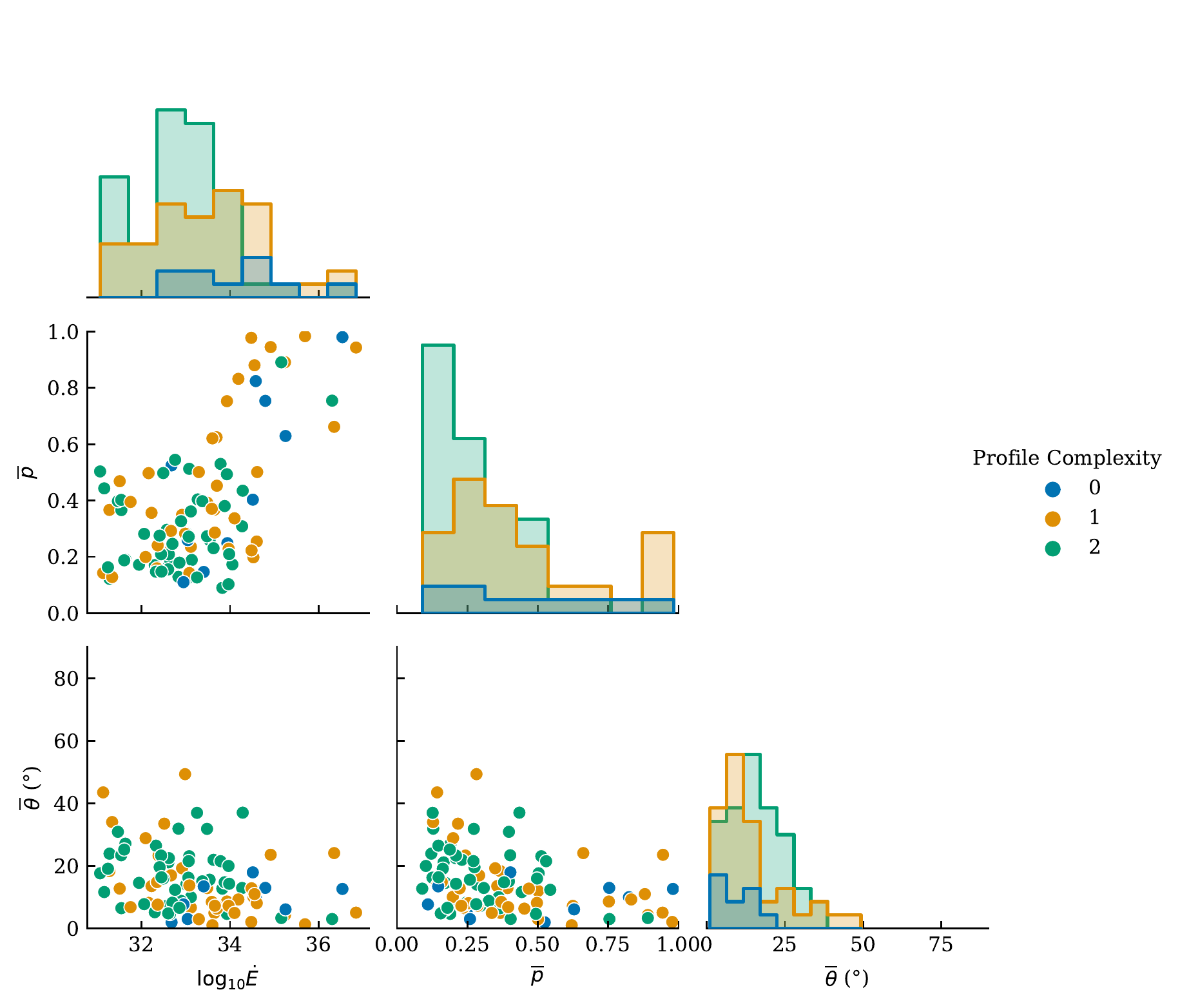}
    \caption{Corner plot of spin-down energy $\dot{E}$, phase-averaged polarization fraction $\overline{p}$ and circular contribution $\overline{\theta}$ (see definitions in Section \ref{sec:data_analysis_methods}) for the 101 high-S/N pulsar subset, split in colour by Profile Complexity category (see definition in Section \ref{subsec:polcategories}).}
    \label{fig:pairplot}
\end{figure*}

Next, we investigate the relationships between polarization measurements and spin-down energy $\dot{E}$ (measured in erg~s$^{-1}$). The association of high $\dot{E}$ pulsars with strong linear polarization has been shown before \citep[e.g.][]{Weltevrede2008}, and it has been noted that high $\dot{E}$ pulsars tend to have simple profiles \citep{Johnston2006}. In addition, pulsars with more complex intensity profile shapes tend to have more complex PA profiles as well: this has been shown for young pulsars \citep{Karastergiou2006} and is particularly well demonstrated by millisecond pulsars \citep{Dai2015}. We investigate these factors together. Fig. \ref{fig:pairplot} shows the distributions of $\dot{E}$, average polarization fraction $\overline{p}$ and circular contribution $\overline{\theta}$ at 1400~MHz, divided into three colours by profile complexity category. Simpler profiles correspond to higher $\dot{E}$, a more uniform distribution of polarization fraction and lower circular contribution, whilst more complex profiles are linked to lower $\dot{E}$, lower polarization fractions and a wider range of $\overline{\theta}$. This work not only replicates the previous results relating high $\dot{E}$ to high linear polarization and simple profile shapes, but demonstrates the associations of all of these factors across $\dot{E}$, and in particular shows that more complex pulsars tend to have a larger proportion of circular polarization (higher $\overline{\theta}$) than simpler profiles. 

We also consider these relations for the whole sample of pulsars, including all of the lower S/N pulsars for which we do not categorize profile complexity. Applying principal component analysis (PCA) to the 3D parameter space of $\log_{10}(\dot{E})$, $\overline{p}$ and $\overline{\theta}$, we find that the principal component vector in this parameter space can be described by the following parametrization of deviations with respect to the mean values: 
\begin{equation}
     \begin{pmatrix}\log_{10}(\dot{E_{0}})\\\overline{p}_{0}\\\overline{\theta}_{0}\end{pmatrix} + t\begin{pmatrix}\Delta\log_{10}(\dot{E})\\\Delta \overline{p}\\\Delta\overline{\theta}\end{pmatrix} = \begin{pmatrix}33.7\\0.4\\13.7$\degree$\end{pmatrix} + t\begin{pmatrix}1.0\\0.2\\-5.1$\degree$\end{pmatrix}.
\end{equation}
This means that an increase in $\dot{E}$ by a factor of 10 corresponds roughly to an increase in polarization fraction of 20\% and a decrease of circular contribution of around 5$\degree$ (i.e. 6\% of 90$\degree$). We note that this is a simplified description which does not encompass the distribution of results around that principle component vector: \cite{Weltevrede2008} described the relationship between $L$ and $\dot{E}$ as showing an arctangent relationship rather than a linear evolution. Our work indicates that it is possible to infer a smoother evolution, though does not confirm absolutely one way or the other. The results of the Thousand-Pulsar-Array census on the MeerKAT telescope will provide a larger population of measurements for investigating this relationship \citep{Posselt2022y}.

\subsection{Combining categorizations and average polarization measurements}

\begin{figure*}
    \centering
    \includegraphics[width=0.8\textwidth]{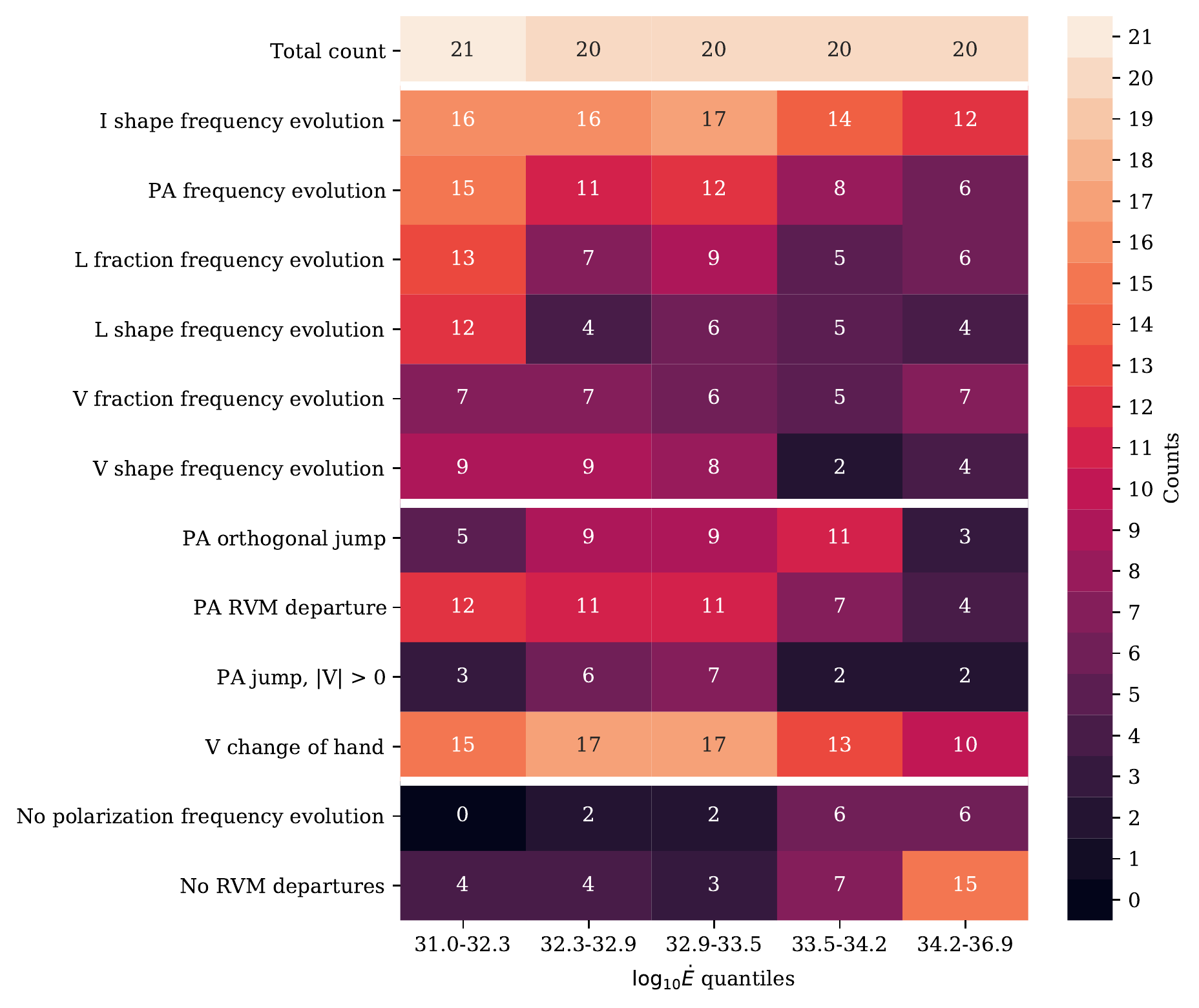}
    \caption{Heat map of the number of counts of pulsars from the high-S/N sample assigned to each polarization category (y-axis), split by their measurement of spin-down energy $\dot{E}$ into five quantiles, such that there is the same number of pulsars in each quantile (x-axis), as shown by the top row which indicates the total count of pulsars in the quantile. A paler colour indicates a higher count, and the actual count in each category is written on the heat map. The polarization categories are defined in Section \ref{subsec:polcategories}.}
    \label{fig:EdotQuantiles}
\end{figure*}

\begin{figure*}
    \centering
    \includegraphics[width=0.8\textwidth]{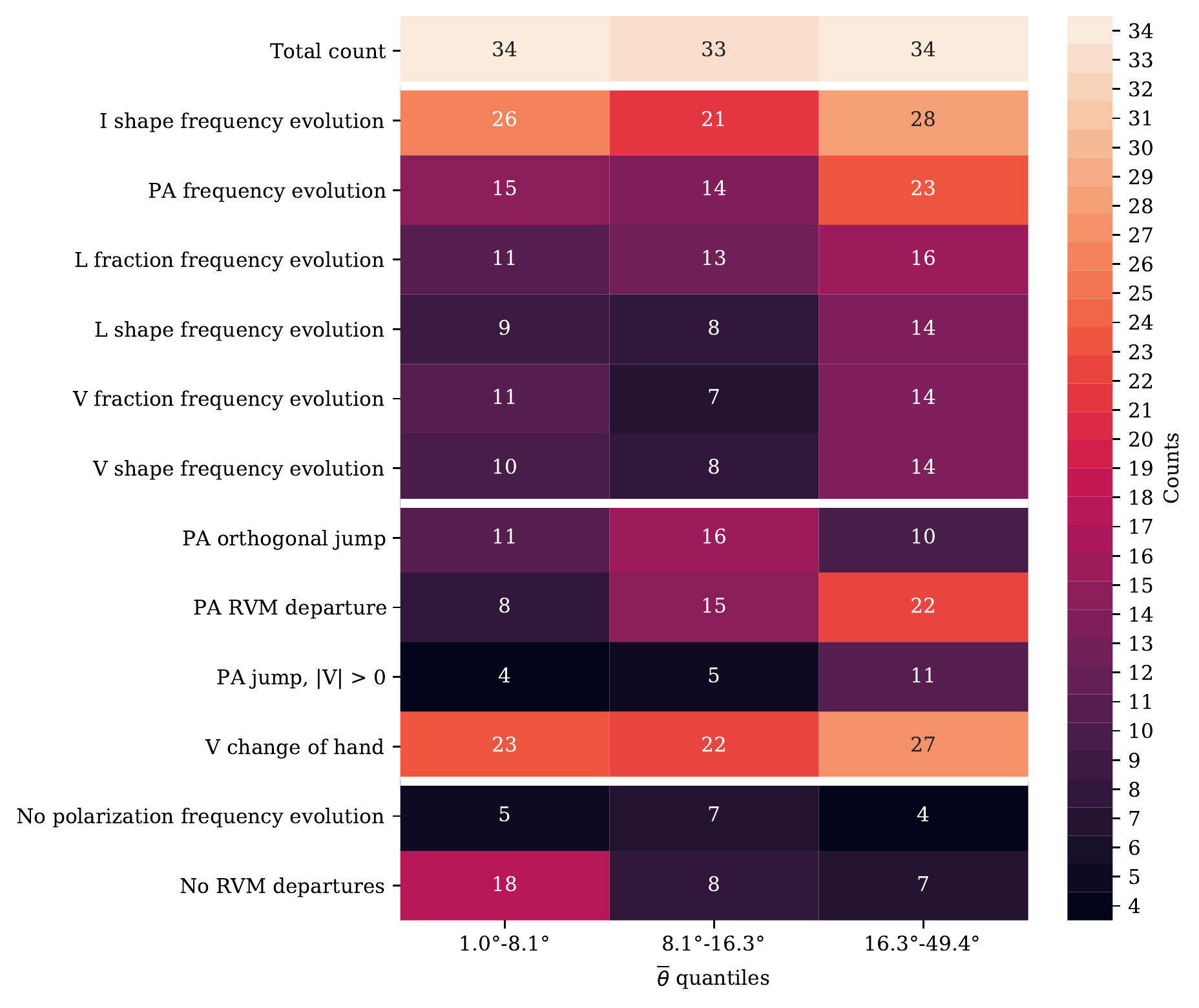}
    \caption{As for Fig.~\ref{fig:EdotQuantiles}, but now divided by three quantiles for circular contribution $\overline{\theta}$ (see definition in Section \ref{sec:data_analysis_methods}). }
    \label{fig:thetaQuantiles}
\end{figure*}

Having considered polarization categories in Section \ref{sec:cat_results} and polarization and $\dot{E}$ measurements in Section \ref{sec:count_results}, we now look to combine the two. Figures \ref{fig:EdotQuantiles} and \ref{fig:thetaQuantiles} compare the 10 polarization categories against $\dot{E}$ and circular contribution $\overline{\theta}$ respectively. For each figure we divide the distribution of $\log_{10}(\dot{E})$ and $\overline{\theta}$ measurements into quantiles such that each quantile contains the same number of pulsars. We note that each quantile spans a different range in $\log_{10}(\dot{E})$ space because the sample is not uniformly distributed across $\log_{10}(\dot{E})$. The same is true for $\bar{\theta}$. We then count how many pulsars in each quantile fall into a given category. On the figures, the categories are divided into frequency-dependent and phase-dependent categories, in a similar way as for Fig. \ref{fig:heatmap}. We also show the total count of pulsars in each quantile, and define two new categories as absences from the other categories: these are simply the counts of pulsars that do not exhibit any frequency evolution of their polarization and do not show departures from the RVM. In Figs \ref{fig:EdotQuantiles} and \ref{fig:thetaQuantiles}, the count is displayed in each quantile-category box and the colour of the box also represents the count. 

These figures demonstrate three factors: the absolute incidences of each categorized feature, the rates of incidence relative to each other, and the relative incidence rates across $\dot{E}$ and $\overline{\theta}$. A general statement to be drawn from Fig. \ref{fig:EdotQuantiles} is the following: there are more occurrences of frequency- and phase-evolution of polarization at lower spin-down energies, particularly the three lowest quantiles spanning the $\dot{E}$ range $10^{31}$ to $10^{33.5}$~erg~s$^{-1}$. Similarly, pulsars showing no polarization frequency evolution and no RVM departures are predominantly found at high spin-down energies, in the top two quantiles of $\dot{E}$ from $10^{33.5}$ to $10^{36.9}$~erg~s$^{-1}$. It should be noted that, since the incidences of the polarization categories are found to be dependent on $\dot{E}$ and our sample has a different distribution of $\dot{E}$ relative to the population as a whole, the exact percentages listed in Table \ref{tab:percentages} are likely to be somewhat different for the pulsar population as a whole. However, the co-occurrence and quantile-based analysis would be directly transferable to a different pulsar sample. In Fig. \ref{fig:thetaQuantiles}, it can be seen that RVM departures and evolution of the PA with frequency are both more common for pulsars with higher values of $\overline{\theta}$, with both of these categories showing increases across the three quantiles. Conversely, pulsars with no RVM departures are predominantly found to have lower values of $\overline{\theta}$. From both figures it can also be seen that the distributions of the ``PA orthogonal jump'' category are skewed in different directions to those of the ``PA RVM departure'' and ``PA jump, $\lvert V\rvert > 0$'' categories: on a population level, true orthogonal jumps are identifiably different from other PA behaviours. 

\subsection{Polarization fraction and circular contribution across the population and across frequency}

Fig. \ref{fig:pthetafreq} shows the distribution of polarization fraction $\overline{p}$ and circular contribution $\overline{\theta}$ as in Fig. \ref{fig:pairplot}, but now for the full sample of 271 pulsars, not just the high-S/N subset. It also shows the measured values at all eight frequency sub-bands, in order to build up a large scale picture of the distributions of these two parameters with respect to each other. For a subset of the pulsars, we plot lines to show the frequency evolution in $\overline{p}$--$\overline{\theta}$ space, and for the rest, we plot the individual values as grey dots. 

\subsubsection{Population distribution}
It can be seen that the distribution is weighted towards the bottom-left of the plot: low polarization and low circular contribution, and that there exist no points at all in the top right quadrant. This means that whereas weakly polarized pulsars have a greater range of circular contribution, strongly polarized pulsars are linearly polarized. This is the same result as shown for the high-S/N subset in Fig. \ref{fig:pairplot}, now shown to be true across the whole observed sample and frequency band.

\subsubsection{Frequency evolution}
Since the polarization measurements presented here are averaged across pulse phase, the values of $\overline{p}$ and $\overline{\theta}$ for a particular pulsar are not fully representative of its polarization behaviour. However, taken collectively, they provide useful insights into the population as a whole. Similarly, for each pulsar, the values for the eight frequency channels form a track in $\overline{p}$--$\overline{\theta}$ space, but individual tracks for each pulsar are expected to be noisy and difficult to interpret. We take an interest therefore in the behaviour of only the straightest tracks in $\overline{p}$--$\overline{\theta}$ space: does their collective behaviour reveal anything interesting about the frequency evolution of pulsar polarization? 

\begin{figure}
    \centering
    \includegraphics[width=\columnwidth]{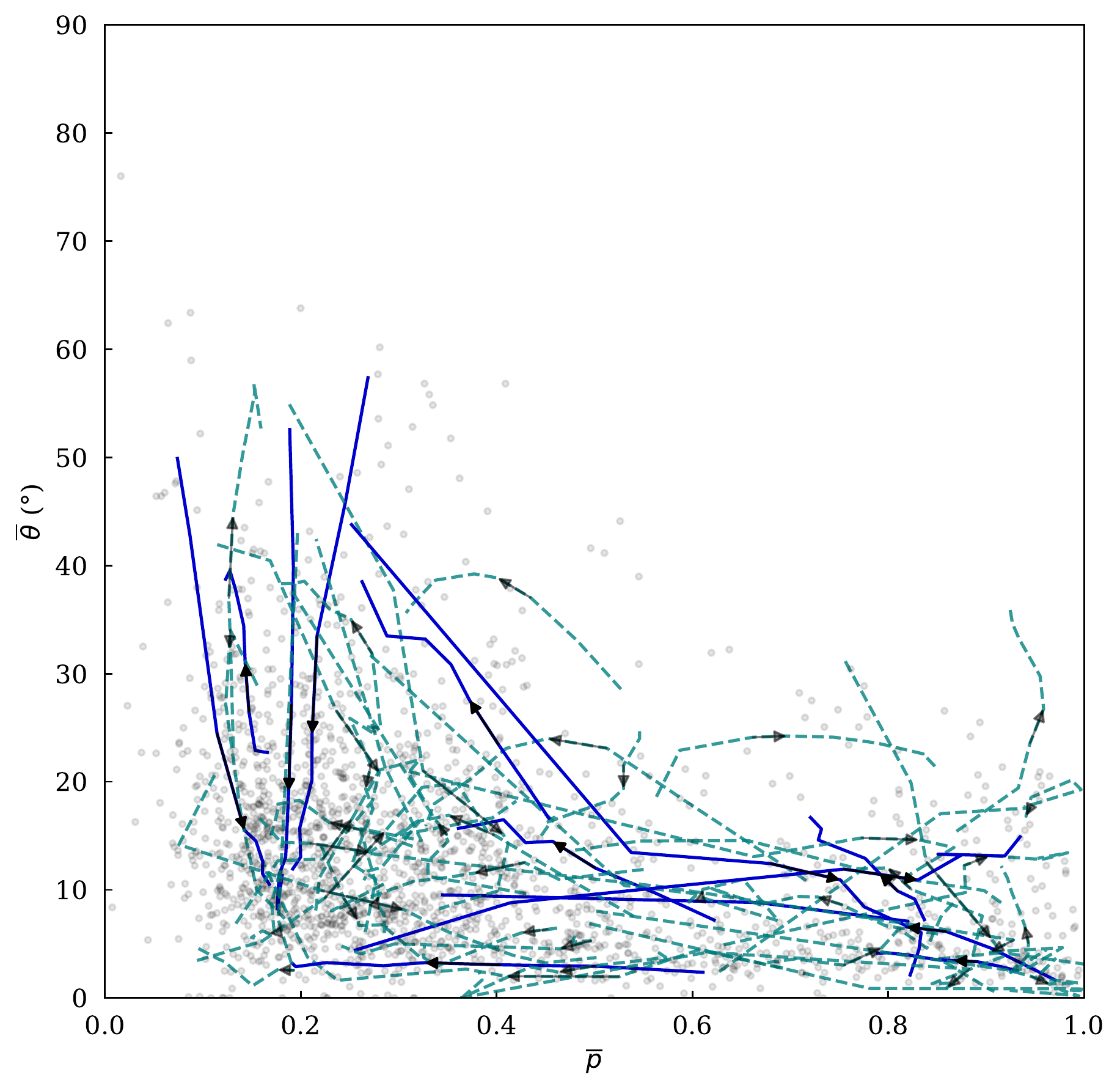}
    \caption{Plot of circular contribution $\overline{\theta}$ vs polarization fraction $\overline{p}$ (see definitions in Section \ref{sec:data_analysis_methods}) for the whole pulsar sample. Pale grey points: individual measurements for each of eight frequency channels for all 271 pulsars, including the measurements for 15 interpulses. Dark and light blue vectors: tracks for individual pulsars showing the evolution of these parameters with increasing frequency. The subset shown have been selected as fitting criteria for straightness of line: the method of selection and justification for this choice are outlined in the text.}
    \label{fig:pthetafreq}
\end{figure}

To identify these, we define two metrics for line length in $\overline{p}$--$\overline{\theta}$ space: the full summed length of the line from one end to the other and its characteristic length, which we define as the furthest distance between two points in the collection of eight values. For a noisy track that walks randomly around a tight area of parameter space, the full summed length of the line will be much longer than the characteristic length, whereas for a perfectly straight line the ratio of the full and characteristic lengths will be 1. We calculate these ratios with $\overline{\theta}$ normalized by 90$\degree$, so that travel through both $\overline{p}$ and $\overline{\theta}$ space is equally weighted. In Fig. \ref{fig:pthetafreq}, we plot only those tracks where the ratio of full-length/characteristic-length < 1.3 (light blue dashed vectors) and full-length/characteristic-length < 1.1 (dark blue vectors). We also plot arrow directions on the vectors to indicate the direction of increasing frequency. 

It can be seen from the figure that at high polarization, the lines plotted tend to be close to horizontal, or following diagonal tracks with a shallow negative gradient, whereas at lower polarization, the gradients of these tracks tend to become steeper. This suggests that the collective $\overline{p}$--$\overline{\theta}$ spread seen as a population-level effect is also born out by the frequency-evolution of individual pulsars. The arrows mainly point in the negative-$\bar{p}$ direction, fitting the general observed trend that pulsars tend to depolarize with increasing frequency. A more in-depth investigation of phase-resolved frequency variation of pulsar polarization would be required to confirm that this effect is truly representative of the whole population, rather than just the least noisy measurements extracted here. To aid in such future investigations, we present waterfall plots of phase- and frequency-resolved polarization fraction $p$ and circular contribution $\theta$ in the second part of the supplementary material (file name ``Appendix C Visualizations of p and theta for pulsar sample'').

\section{Discussion}
\label{sec:summary}

\subsection{Summary of key results}

A large fraction of pulsars show frequency and phase evolution not encompassed by the conventional picture \citep{lorimer2005handbook}. We find links between non-RVM behaviour of the PA, frequency evolution of polarization and the presence of circular polarization features. Non-RVM behaviour of the PA, as defined by categories (ii)b and (ii)c, is distinct from the behaviour of true orthogonal jumps.

In general, pulsars with higher $\dot{E}$ also have simpler pulse profiles, higher polarization fractions and lower values of $\overline{\theta}$. There exist no high-$\overline{p}$-high-$\overline{\theta}$ pulsars. Frequency evolution of polarization is seen more in low $\dot{E}$ pulsars. Similarly, phase evolution of polarization, particularly non-RVM behaviour of the PA, is observed more frequently in low $\dot{E}$ pulsars, whereas high $\dot{E}$ pulsars demonstrate little to no frequency evolution of polarization, along with RVM-like PA profiles. 

Polarization measurements evolve with frequency on curved tracks from high $\overline{p}$, low $\overline{\theta}$ to low $\overline{p}$, high $\overline{\theta}$ for at least some pulsars and for average profile observations. In addition, lower $\overline{p}$ corresponds to steeper tracks: more weakly polarized pulsars have stronger frequency evolution of circular polarization.

\subsection{Interpretation of results}

An explanation previously presented for the links between profile complexity and $\dot{E}$ is that older and younger pulsars (with lower and higher values of $\dot{E}$) may have different emission height profiles \citep{Karastergiou2007}. A narrower range of emission heights for high $\dot{E}$ pulsars could result in simpler profiles, whilst a broader range of emission heights for low $\dot{E}$ pulsars might introduce more complex profile shapes and with them the potential for increased frequency evolution, non-RVM behaviour and reduced polarization fraction. 

Under such a scenario, the links between circular polarization features, non-RVM behaviour in the PA and frequency evolution of polarization could be explained as originating as propagational effects in the magnetosphere. Propagational effects discussed in the literature include angular separation of orthogonal modes due to refraction \citep{Barnard1986}, Generalized Faraday Rotation \citep{Kennett1998} or coherent mixing of modes \citep{Dyks2017d}. The broad bandwidth of the UWL highlights the relevance of frequency-dependent effects in the polarization, which in future could be probed further across even wider bandwidths: current suitable low-frequency instruments include LOFAR and the MWA, and higher frequency observations will be possible with future development of the MeerKAT S-band and Parkes Ultra-Wideband High receivers. In the future, the Square Kilometre Array will increase the sensitivity of pulsar observations, increasing the available high-S/N sample size for detailed analysis.

The approach taken in this paper is to consider the collective polarization behaviour of the  pulsar population, reducing the degrees of freedom of the dataset by calculating average polarizations and defining boolean categories to describe the features and trends of the dataset. It can immediately be seen from the online supplementary material that each individual pulsar also displays its own idiosyncratic and interesting polarization behaviour. This study focuses on young, non-recycled pulsars, but the polarization idiosyncrasies of millisecond pulsars \citep[e.g.][]{Dai2015} and magnetars \citep[e.g.][]{Dai2019a} are observed to be even more extreme. Descriptions of radio pulsar polarization must ultimately be able to account for the diversity of individual polarization behaviours observed in broad-band pulsar data whilst also explaining the origins of the patterns observed across the pulsar population, as described here. 

We propose a model that can achieve this: the ``partially-coherent'' model of orthogonal mode mixing. In a follow-up work, we show how this model can explain the relationship between polarization fraction and circular contribution and the link between non-RVM PA behaviour and the presence of circular polarization. The partially-coherent model can account for the polarization trends described in this paper and can be used as a mathematical decomposition of Stokes parameters to account for the idiosyncrasies of individual pulsars as well.

\section{Conclusions}
\label{sec:conc}

The broad-band capabilities of the Parkes UWL observing system, and the long-term observations of this dataset, provide an opportunity for an updated picture of the polarization properties of young pulsars. This work presents broad-band phase-resolved polarization information for 271 pulsars with a new waterfall plot visualization, to provide a basis for a new description of radio pulsar polarization. By defining and comparing collective categories for polarization behaviour for the 101 pulsars passing our S/N criterion, we demonstrate the relationships between key features of pulsar profiles, encompassing links between phase- and frequency-dependent polarization behaviour. We highlight the importance of circular polarization as a key contributing factor to non-RVM polarization behaviour, and an important measurable quantity for understanding the origins of pulsar radio polarization. It is increasingly clear that physical models of pulsar radio polarization must be able to explain the complexities of broad-band pulse profile evolution and the key contribution of circular polarization. In follow-up work, we outline how we can do this using the partially-coherent model of orthogonal mode interaction. 

\section*{Acknowledgements}
We would like to thank the reviewer for useful advice on how to improve the quality of the manuscript. The Parkes radio telescope (Murriyang) is part of the Australia Telescope National Facility (\href{https://ror.org/05qajvd42}{https://ror.org/05qajvd42}) which is funded by the Australian Government for operation as a National Facility managed by CSIRO. 
We acknowledge the Wiradjuri people as the traditional owners of the Observatory site. 
LSO acknowledges the support of Magdalen College, Oxford. 
SD is the recipient of an Australian Research Council Discovery Early Career Award (DE210101738) funded by the Australian Government. 
Work at NRL is supported by NASA. 
RMS acknowledges support through Australian Research Council Future Fellowship FT190100155. 

\section*{Data Availability}
The data underlying this article will be shared on reasonable request to the corresponding author.

%%%%%%%%%%%%%%%%%%%%%%%%%%%%%%%%%%%%%%%%%%%%%%%%%%



\bibliographystyle{mnras}
\bibliography{CircularPolarization} % if your bibtex file is called example.bib

%%%%%%%%%%%%%%%%%%%%%%%%%%%%%%%%%%%%%%%%%%%%%%%%%%

%%%%%%%%%%%%%%%%% APPENDICES %%%%%%%%%%%%%%%%%%%%%
\section*{Supplementary material}
Visualizations of all of the pulsars in this sample are available at MNRAS online. The two pieces of supplementary material are labelled ``Appendix B Polarization visualizations for pulsar sample'' and ``Appendix C Visualizations of p and theta for pulsar sample''.

\appendix

\section{Table of polarization measurements and categorizations}

\onecolumn
\begin{longtable}{llllllll}
\caption{Results of polarization measurements and categorizations for all 271 pulsars (plus 15 interpulses) in the sample. Column labels are defined as follows: `PSRJ' is the pulsar name. Interpulses are indicated with `(i)'. $\overline{l}$, $\overline{\lvert v\rvert}$, $\overline{p}$ and $\overline{\theta}$ are polarization fractions and circular contribution as defined in Section \ref{sec:data_analysis_methods}. `Passed S/N cut' is marked with `Y' if the pulsar is part of the high-S/N subset as defined in Section \ref{sec:data_analysis_methods}. Column $S$ indicates scattered pulsars with an asterisk. `Categories' lists the polarization categorization applied to the high-S/N pulsar sample as described in Section \ref{subsec:polcategories}.}\label{tab:all_ave_psr_results}\\
\toprule
PSRJ & $\overline{l}$ & $\overline{\lvert v\rvert}$ & $\overline{p}$ & $\overline{\theta}$    & Passed  & $S$ & Categories \\
     &      &                   &      & ($\degree$)  & S/N cut &     &            \\
\midrule
\endfirsthead
Table A1 (continued) \\
\midrule
PSRJ & $\overline{l}$ & $\overline{\lvert v\rvert}$ & $\overline{p}$ & $\overline{\theta}$    & Passed  & $S$ & Categories \\
     &      &                   &      & ($\degree$)  & S/N cut &     &            \\
\midrule
\endhead
\midrule
\multicolumn{8}{r}{{Continued on next page}} \\
\midrule
\endfoot

\bottomrule
\endlastfoot
     J0034-0721 &      0.100 $\pm$ 0.002 &      0.044 $\pm$ 0.002 &      0.122 $\pm$ 0.002 &             24 $\pm$ 1 &              Y &     &      (i)abcd;(ii)a;2 \\
     J0108-1431 &      0.768 $\pm$ 0.009 &      0.063 $\pm$ 0.007 &      0.781 $\pm$ 0.009 &          4.7 $\pm$ 0.5 &                &     &                      \\
     J0134-2937 &      0.454 $\pm$ 0.003 &      0.194 $\pm$ 0.003 &      0.513 $\pm$ 0.003 &         23.1 $\pm$ 0.4 &              Y &     &       (i)abd;(ii)a;2 \\
     J0151-0635 &      0.324 $\pm$ 0.005 &      0.041 $\pm$ 0.005 &      0.338 $\pm$ 0.005 &          7.2 $\pm$ 0.9 &                &     &                      \\
     J0152-1637 &      0.154 $\pm$ 0.006 &      0.040 $\pm$ 0.006 &      0.172 $\pm$ 0.006 &             15 $\pm$ 2 &              Y &     &      (i)abd;(ii)cd;2 \\
     J0206-4028 &      0.137 $\pm$ 0.006 &      0.070 $\pm$ 0.006 &      0.179 $\pm$ 0.006 &             27 $\pm$ 4 &                &     &                      \\
     J0255-5304 &      0.088 $\pm$ 0.002 &      0.084 $\pm$ 0.002 &      0.142 $\pm$ 0.002 &             44 $\pm$ 3 &              Y &     &        (i)b;(ii)cd;1 \\
     J0304+1932 &      0.334 $\pm$ 0.003 &      0.111 $\pm$ 0.003 &      0.366 $\pm$ 0.003 &         18.4 $\pm$ 0.6 &              Y &     &     (i)abcdf;(ii)d;1 \\
     J0401-7608 &      0.283 $\pm$ 0.006 &      0.036 $\pm$ 0.005 &      0.296 $\pm$ 0.006 &              7 $\pm$ 1 &              Y &     &    (i)abd;(ii)abcd;2 \\
     J0448-2749 &      0.244 $\pm$ 0.005 &      0.138 $\pm$ 0.005 &      0.295 $\pm$ 0.005 &             30 $\pm$ 2 &                &     &                      \\
     J0452-1759 &    0.1850 $\pm$ 0.0005 &    0.0216 $\pm$ 0.0004 &    0.1887 $\pm$ 0.0005 &          6.7 $\pm$ 0.1 &              Y &     &    (i)abdf;(ii)abd;2 \\
     J0525+1115 &      0.127 $\pm$ 0.006 &      0.119 $\pm$ 0.006 &      0.209 $\pm$ 0.006 &             43 $\pm$ 7 &                &     &                      \\
     J0536-7543 &      0.466 $\pm$ 0.002 &      0.148 $\pm$ 0.002 &      0.503 $\pm$ 0.002 &         17.7 $\pm$ 0.3 &              Y &     &       (i)cdef;(ii);2 \\
     J0543+2329 &      0.483 $\pm$ 0.001 &      0.103 $\pm$ 0.001 &      0.501 $\pm$ 0.001 &         12.0 $\pm$ 0.2 &              Y &     &          (i)c;(ii);1 \\
     J0601-0527 &      0.309 $\pm$ 0.004 &      0.108 $\pm$ 0.004 &      0.349 $\pm$ 0.004 &         19.3 $\pm$ 0.9 &              Y &     &      (i)af;(ii)acd;1 \\
     J0614+2229 &      0.723 $\pm$ 0.003 &      0.167 $\pm$ 0.003 &      0.754 $\pm$ 0.003 &         13.0 $\pm$ 0.2 &              Y &     &           (i);(ii);0 \\
     J0624-0424 &      0.219 $\pm$ 0.007 &      0.063 $\pm$ 0.007 &      0.267 $\pm$ 0.007 &             16 $\pm$ 2 &                &     &                      \\
     J0627+0706 &      0.233 $\pm$ 0.005 &      0.038 $\pm$ 0.005 &      0.244 $\pm$ 0.005 &              9 $\pm$ 1 &                &     &                      \\
 J0627+0706 (i) &        0.17 $\pm$ 0.01 &        0.05 $\pm$ 0.01 &        0.23 $\pm$ 0.01 &             18 $\pm$ 5 &                &     &                      \\
     J0630-2834 &    0.4901 $\pm$ 0.0007 &    0.0706 $\pm$ 0.0006 &    0.4971 $\pm$ 0.0007 &        8.19 $\pm$ 0.08 &              Y &     &          (i)d;(ii);1 \\
     J0631+1036 &        0.87 $\pm$ 0.01 &      0.066 $\pm$ 0.009 &        0.89 $\pm$ 0.01 &          4.3 $\pm$ 0.6 &              Y &     &         (i)a;(ii)d;1 \\
     J0659+1414 &      0.797 $\pm$ 0.005 &      0.141 $\pm$ 0.004 &      0.824 $\pm$ 0.005 &         10.0 $\pm$ 0.3 &              Y &     &         (i)ce;(ii);0 \\
     J0729-1448 &        0.89 $\pm$ 0.01 &        0.11 $\pm$ 0.01 &        0.91 $\pm$ 0.01 &          7.0 $\pm$ 0.6 &                &     &                      \\
     J0729-1836 &      0.266 $\pm$ 0.005 &      0.063 $\pm$ 0.005 &      0.282 $\pm$ 0.005 &             13 $\pm$ 1 &                &     &                      \\
     J0738-4042 &    0.2743 $\pm$ 0.0001 &    0.0686 $\pm$ 0.0001 &    0.2838 $\pm$ 0.0001 &       14.04 $\pm$ 0.03 &              Y &     &     (i)abcd;(ii)ad;2 \\
     J0742-2822 &    0.8888 $\pm$ 0.0005 &    0.0521 $\pm$ 0.0003 &    0.8909 $\pm$ 0.0005 &        3.35 $\pm$ 0.02 &              Y &     &        (i)ac;(ii)d;2 \\
     J0745-5353 &      0.240 $\pm$ 0.003 &      0.025 $\pm$ 0.003 &      0.248 $\pm$ 0.003 &          5.9 $\pm$ 0.6 &              Y &     &         (i)d;(ii)d;0 \\
     J0758-1528 &      0.168 $\pm$ 0.003 &      0.015 $\pm$ 0.003 &      0.170 $\pm$ 0.003 &              5 $\pm$ 1 &              Y &     &       (i)ab;(ii)cd;2 \\
     J0809-4753 &      0.275 $\pm$ 0.003 &      0.040 $\pm$ 0.003 &      0.286 $\pm$ 0.003 &          8.3 $\pm$ 0.7 &                &     &                      \\
     J0820-1350 &      0.162 $\pm$ 0.001 &      0.083 $\pm$ 0.001 &      0.190 $\pm$ 0.001 &         27.1 $\pm$ 0.5 &              Y &     &   (i)abcdef;(ii)cd;2 \\
     J0820-3826 &        0.24 $\pm$ 0.02 &        0.04 $\pm$ 0.02 &        0.30 $\pm$ 0.02 &              9 $\pm$ 5 &                &     &                      \\
     J0820-4114 &      0.279 $\pm$ 0.003 &      0.091 $\pm$ 0.003 &      0.313 $\pm$ 0.003 &         18.0 $\pm$ 0.6 &                &     &                      \\
 J0834-4159 (i) &          0.2 $\pm$ 0.2 &          0.0 $\pm$ 0.2 &          0.2 $\pm$ 0.2 &             0 $\pm$ 57 &                &     &                      \\
     J0834-4159 &        0.12 $\pm$ 0.03 &        0.13 $\pm$ 0.03 &        0.31 $\pm$ 0.03 &            47 $\pm$ 49 &                &     &                      \\
     J0835-4510 &  0.93794 $\pm$ 0.00007 &  0.08339 $\pm$ 0.00005 &  0.94321 $\pm$ 0.00007 &      5.081 $\pm$ 0.003 &              Y &     &       (i)acde;(ii);1 \\
     J0837+0610 &      0.093 $\pm$ 0.002 &      0.051 $\pm$ 0.002 &      0.115 $\pm$ 0.002 &             29 $\pm$ 2 &                &     &                      \\
     J0837-4135 &    0.1685 $\pm$ 0.0006 &    0.1115 $\pm$ 0.0006 &    0.2166 $\pm$ 0.0006 &         33.5 $\pm$ 0.3 &              Y &     &    (i)abef;(ii)acd;1 \\
 J0842-4851 (i) &        0.07 $\pm$ 0.04 &        0.02 $\pm$ 0.04 &        0.10 $\pm$ 0.04 &            13 $\pm$ 32 &                &     &                      \\
     J0842-4851 &      0.083 $\pm$ 0.008 &      0.014 $\pm$ 0.008 &      0.100 $\pm$ 0.008 &             10 $\pm$ 6 &                &     &                      \\
     J0855-4644 &        0.54 $\pm$ 0.03 &        0.05 $\pm$ 0.02 &        0.60 $\pm$ 0.03 &              6 $\pm$ 3 &                &     &                      \\
     J0857-4424 &        0.08 $\pm$ 0.01 &        0.04 $\pm$ 0.01 &        0.12 $\pm$ 0.01 &             25 $\pm$ 9 &                &     &                      \\
     J0901-4624 &        0.60 $\pm$ 0.02 &        0.30 $\pm$ 0.02 &        0.75 $\pm$ 0.02 &             27 $\pm$ 3 &                &     &                      \\
     J0904-7459 &      0.245 $\pm$ 0.009 &      0.013 $\pm$ 0.009 &      0.257 $\pm$ 0.009 &              3 $\pm$ 2 &                &     &                      \\
 J0905-5127 (i) &          0.2 $\pm$ 0.1 &          0.0 $\pm$ 0.1 &          0.2 $\pm$ 0.1 &             7 $\pm$ 38 &                &     &                      \\
     J0905-5127 &        0.84 $\pm$ 0.01 &      0.091 $\pm$ 0.009 &        0.86 $\pm$ 0.01 &          6.2 $\pm$ 0.7 &                &     &                      \\
     J0907-5157 &      0.359 $\pm$ 0.001 &      0.031 $\pm$ 0.001 &      0.367 $\pm$ 0.001 &          5.0 $\pm$ 0.2 &              Y &     &      (i)abdf;(ii)d;1 \\
     J0908-1739 &      0.140 $\pm$ 0.003 &      0.054 $\pm$ 0.003 &      0.166 $\pm$ 0.003 &             21 $\pm$ 1 &              Y &     &      (i)abd;(ii)cd;2 \\
 J0908-4913 (i) &      0.966 $\pm$ 0.005 &      0.035 $\pm$ 0.004 &      0.969 $\pm$ 0.005 &          2.1 $\pm$ 0.2 &              Y &     &       (i)aef;(ii)d;1 \\
     J0908-4913 &      0.982 $\pm$ 0.002 &      0.023 $\pm$ 0.001 &      0.984 $\pm$ 0.002 &        1.35 $\pm$ 0.07 &              Y &     &       (i)aef;(ii)d;1 \\
     J0924-5814 &      0.521 $\pm$ 0.004 &      0.018 $\pm$ 0.003 &      0.525 $\pm$ 0.004 &          2.0 $\pm$ 0.4 &              Y &     &          (i)c;(ii);0 \\
     J0940-5428 &        0.60 $\pm$ 0.02 &        0.04 $\pm$ 0.02 &        0.63 $\pm$ 0.02 &              4 $\pm$ 2 &                &     &                      \\
     J0942-5552 &      0.372 $\pm$ 0.001 &      0.085 $\pm$ 0.001 &      0.392 $\pm$ 0.001 &         12.9 $\pm$ 0.2 &              Y &     &       (i)a;(ii)acd;1 \\
     J0954-5430 &        0.55 $\pm$ 0.02 &        0.08 $\pm$ 0.01 &        0.57 $\pm$ 0.02 &              9 $\pm$ 2 &                &     &                      \\
     J0959-4809 &        0.38 $\pm$ 0.01 &      0.030 $\pm$ 0.009 &        0.40 $\pm$ 0.01 &              5 $\pm$ 1 &                &     &                      \\
     J1001-5507 &      0.100 $\pm$ 0.001 &      0.062 $\pm$ 0.001 &      0.129 $\pm$ 0.001 &             32 $\pm$ 1 &              Y &     &    (i)abde;(ii)acd;2 \\
     J1003-4747 &      0.151 $\pm$ 0.008 &      0.023 $\pm$ 0.008 &      0.177 $\pm$ 0.008 &              9 $\pm$ 3 &                &     &                      \\
     J1012-5857 &      0.130 $\pm$ 0.005 &      0.032 $\pm$ 0.005 &      0.144 $\pm$ 0.005 &             14 $\pm$ 2 &              Y &   * &                      \\
     J1015-5719 &        0.80 $\pm$ 0.01 &      0.108 $\pm$ 0.008 &        0.83 $\pm$ 0.01 &          7.7 $\pm$ 0.6 &                &     &                      \\
     J1016-5819 &        0.17 $\pm$ 0.02 &        0.05 $\pm$ 0.02 &        0.22 $\pm$ 0.03 &            17 $\pm$ 10 &                &     &                      \\
     J1016-5857 &        0.50 $\pm$ 0.02 &        0.08 $\pm$ 0.02 &        0.57 $\pm$ 0.03 &              9 $\pm$ 3 &                &     &                      \\
     J1017-5621 &      0.168 $\pm$ 0.004 &      0.196 $\pm$ 0.004 &      0.282 $\pm$ 0.004 &             49 $\pm$ 5 &              Y &     &      (i)bce;(ii)cd;1 \\
     J1019-5749 &        0.17 $\pm$ 0.01 &        0.01 $\pm$ 0.01 &        0.21 $\pm$ 0.01 &              5 $\pm$ 4 &              Y &   * &                      \\
     J1028-5819 &        0.99 $\pm$ 0.02 &        0.01 $\pm$ 0.02 &        0.99 $\pm$ 0.02 &          0.8 $\pm$ 0.9 &                &     &                      \\
     J1034-3224 &      0.228 $\pm$ 0.004 &      0.039 $\pm$ 0.004 &      0.253 $\pm$ 0.004 &          9.7 $\pm$ 0.9 &                &     &                      \\
     J1038-5831 &      0.333 $\pm$ 0.008 &      0.080 $\pm$ 0.007 &      0.356 $\pm$ 0.008 &             14 $\pm$ 1 &              Y &     &     (i)acef;(ii)ad;1 \\
     J1043-6116 &      0.130 $\pm$ 0.006 &      0.111 $\pm$ 0.006 &      0.208 $\pm$ 0.006 &             41 $\pm$ 6 &              Y &   * &                      \\
     J1046-5813 &      0.145 $\pm$ 0.007 &      0.034 $\pm$ 0.006 &      0.167 $\pm$ 0.007 &             13 $\pm$ 3 &                &     &                      \\
     J1047-6709 &      0.737 $\pm$ 0.005 &      0.112 $\pm$ 0.004 &      0.753 $\pm$ 0.005 &          8.6 $\pm$ 0.4 &              Y &     &           (i);(ii);1 \\
     J1048-5832 &      0.751 $\pm$ 0.001 &      0.040 $\pm$ 0.001 &      0.755 $\pm$ 0.001 &        3.03 $\pm$ 0.08 &              Y &     &        (i)ae;(ii)d;2 \\
     J1049-5833 &        0.23 $\pm$ 0.01 &        0.04 $\pm$ 0.01 &        0.26 $\pm$ 0.01 &              9 $\pm$ 3 &                &     &                      \\
     J1055-6028 &        0.26 $\pm$ 0.01 &        0.03 $\pm$ 0.01 &        0.28 $\pm$ 0.01 &              6 $\pm$ 2 &                &     &                      \\
     J1056-6258 &    0.4021 $\pm$ 0.0006 &    0.0215 $\pm$ 0.0005 &    0.4037 $\pm$ 0.0006 &        3.07 $\pm$ 0.08 &              Y &     &         (i)ac;(ii);2 \\
     J1057-5226 &      0.975 $\pm$ 0.005 &      0.036 $\pm$ 0.003 &      0.978 $\pm$ 0.005 &          2.1 $\pm$ 0.2 &              Y &     &          (i);(ii)d;1 \\
 J1057-5226 (i) &      0.438 $\pm$ 0.004 &      0.023 $\pm$ 0.004 &      0.447 $\pm$ 0.004 &          3.0 $\pm$ 0.5 &              Y &     &          (i);(ii)d;1 \\
     J1105-6107 &        0.89 $\pm$ 0.01 &        0.08 $\pm$ 0.01 &        0.91 $\pm$ 0.01 &          5.4 $\pm$ 0.7 &                &     &                      \\
     J1110-5637 &      0.264 $\pm$ 0.003 &      0.081 $\pm$ 0.003 &      0.292 $\pm$ 0.003 &         17.0 $\pm$ 0.8 &              Y &     &        (i)c;(ii)ad;1 \\
     J1112-6103 &        0.13 $\pm$ 0.01 &        0.02 $\pm$ 0.01 &        0.16 $\pm$ 0.01 &              7 $\pm$ 5 &              Y &   * &                      \\
     J1114-6100 &      0.189 $\pm$ 0.004 &      0.008 $\pm$ 0.004 &      0.193 $\pm$ 0.004 &              3 $\pm$ 1 &                &   * &                      \\
     J1115-6052 &        0.69 $\pm$ 0.02 &        0.05 $\pm$ 0.02 &        0.71 $\pm$ 0.02 &              4 $\pm$ 1 &                &     &                      \\
     J1119-6127 &        0.94 $\pm$ 0.02 &        0.06 $\pm$ 0.01 &        0.95 $\pm$ 0.02 &          3.4 $\pm$ 0.7 &                &     &                      \\
     J1123-6259 &        0.55 $\pm$ 0.01 &        0.04 $\pm$ 0.01 &        0.56 $\pm$ 0.01 &              4 $\pm$ 1 &                &     &                      \\
     J1136-5525 &      0.079 $\pm$ 0.002 &      0.018 $\pm$ 0.002 &      0.090 $\pm$ 0.002 &             13 $\pm$ 2 &              Y &     &        (i)ac;(ii)a;2 \\
     J1146-6030 &      0.242 $\pm$ 0.002 &      0.068 $\pm$ 0.002 &      0.259 $\pm$ 0.002 &         15.6 $\pm$ 0.6 &              Y &     &     (i)abcd;(ii)ad;2 \\
     J1156-5707 &        0.51 $\pm$ 0.02 &        0.04 $\pm$ 0.02 &        0.53 $\pm$ 0.02 &              5 $\pm$ 2 &                &     &                      \\
     J1157-6224 &      0.312 $\pm$ 0.002 &      0.121 $\pm$ 0.002 &      0.373 $\pm$ 0.002 &         21.3 $\pm$ 0.5 &                &     &                      \\
     J1210-5559 &      0.226 $\pm$ 0.005 &      0.026 $\pm$ 0.005 &      0.236 $\pm$ 0.005 &              7 $\pm$ 1 &              Y &     &         (i);(ii)ad;1 \\
     J1224-6407 &      0.324 $\pm$ 0.001 &      0.245 $\pm$ 0.001 &      0.435 $\pm$ 0.001 &         37.0 $\pm$ 0.3 &              Y &     &  (i)abcdef;(ii)bcd;2 \\
     J1225-6408 &      0.335 $\pm$ 0.008 &      0.046 $\pm$ 0.008 &      0.348 $\pm$ 0.008 &              8 $\pm$ 1 &                &     &                      \\
     J1243-6423 &    0.2245 $\pm$ 0.0004 &    0.1394 $\pm$ 0.0004 &    0.2729 $\pm$ 0.0004 &         31.8 $\pm$ 0.2 &              Y &     &    (i)acdef;(ii)cd;2 \\
     J1253-5820 &      0.616 $\pm$ 0.003 &      0.078 $\pm$ 0.003 &      0.625 $\pm$ 0.003 &          7.2 $\pm$ 0.2 &              Y &     &          (i)a;(ii);1 \\
     J1301-6305 &        0.67 $\pm$ 0.04 &        0.05 $\pm$ 0.03 &        0.72 $\pm$ 0.04 &              4 $\pm$ 3 &                &     &                      \\
 J1302-6350 (i) &        0.83 $\pm$ 0.01 &      0.181 $\pm$ 0.009 &        0.88 $\pm$ 0.01 &         12.3 $\pm$ 0.7 &                &     &                      \\
     J1302-6350 &      0.916 $\pm$ 0.009 &      0.055 $\pm$ 0.007 &      0.925 $\pm$ 0.009 &          3.5 $\pm$ 0.4 &                &     &                      \\
     J1305-6203 &        0.44 $\pm$ 0.02 &        0.03 $\pm$ 0.01 &        0.45 $\pm$ 0.02 &              4 $\pm$ 2 &                &     &                      \\
     J1306-6617 &      0.204 $\pm$ 0.003 &      0.060 $\pm$ 0.003 &      0.227 $\pm$ 0.003 &         16.3 $\pm$ 0.9 &              Y &   * &                      \\
     J1317-6302 &      0.114 $\pm$ 0.007 &      0.045 $\pm$ 0.007 &      0.156 $\pm$ 0.007 &             21 $\pm$ 4 &                &   * &                      \\
     J1319-6056 &      0.360 $\pm$ 0.006 &      0.092 $\pm$ 0.006 &      0.387 $\pm$ 0.006 &             14 $\pm$ 1 &                &   * &                      \\
     J1320-5359 &      0.216 $\pm$ 0.005 &      0.066 $\pm$ 0.004 &      0.238 $\pm$ 0.005 &             17 $\pm$ 1 &                &     &                      \\
     J1326-5859 &    0.3412 $\pm$ 0.0004 &    0.1855 $\pm$ 0.0004 &    0.4061 $\pm$ 0.0004 &       28.53 $\pm$ 0.09 &              Y &   * &                      \\
     J1326-6408 &      0.197 $\pm$ 0.005 &      0.085 $\pm$ 0.005 &      0.243 $\pm$ 0.005 &             23 $\pm$ 2 &              Y &     &        (i)a;(ii)cd;1 \\
     J1326-6700 &    0.3524 $\pm$ 0.0008 &    0.0623 $\pm$ 0.0007 &    0.3610 $\pm$ 0.0008 &         10.0 $\pm$ 0.1 &              Y &     &     (i)acd;(ii)acd;2 \\
     J1327-6222 &    0.0990 $\pm$ 0.0005 &    0.0450 $\pm$ 0.0005 &    0.1145 $\pm$ 0.0005 &         24.4 $\pm$ 0.4 &              Y &   * &                      \\
     J1327-6301 &      0.291 $\pm$ 0.004 &      0.022 $\pm$ 0.004 &      0.303 $\pm$ 0.004 &          4.4 $\pm$ 0.7 &                &     &                      \\
     J1328-4357 &      0.316 $\pm$ 0.002 &      0.049 $\pm$ 0.002 &      0.326 $\pm$ 0.002 &          8.8 $\pm$ 0.4 &              Y &     &      (i)ace;(ii)ad;2 \\
     J1338-6204 &      0.270 $\pm$ 0.003 &      0.104 $\pm$ 0.003 &      0.325 $\pm$ 0.003 &         21.0 $\pm$ 0.8 &              Y &   * &                      \\
     J1340-6456 &      0.090 $\pm$ 0.008 &      0.037 $\pm$ 0.008 &      0.123 $\pm$ 0.008 &             22 $\pm$ 6 &                &     &                      \\
     J1341-6220 &      0.643 $\pm$ 0.008 &      0.142 $\pm$ 0.007 &      0.680 $\pm$ 0.008 &         12.4 $\pm$ 0.6 &                &   * &                      \\
     J1349-6130 &        0.28 $\pm$ 0.01 &        0.03 $\pm$ 0.01 &        0.30 $\pm$ 0.01 &              5 $\pm$ 3 &                &     &                      \\
     J1352-6803 &        0.35 $\pm$ 0.02 &        0.06 $\pm$ 0.02 &        0.38 $\pm$ 0.02 &             10 $\pm$ 3 &                &     &                      \\
     J1356-5521 &      0.269 $\pm$ 0.006 &      0.023 $\pm$ 0.006 &      0.284 $\pm$ 0.006 &              5 $\pm$ 1 &                &     &                      \\
       J1357-62 &      0.199 $\pm$ 0.001 &      0.100 $\pm$ 0.001 &      0.236 $\pm$ 0.001 &         26.6 $\pm$ 0.5 &              Y &   * &                      \\
     J1357-6429 &        0.82 $\pm$ 0.03 &        0.09 $\pm$ 0.02 &        0.86 $\pm$ 0.03 &              6 $\pm$ 2 &                &     &                      \\
     J1359-6038 &      0.789 $\pm$ 0.001 &      0.138 $\pm$ 0.001 &      0.810 $\pm$ 0.001 &        9.90 $\pm$ 0.07 &              Y &   * &                      \\
     J1401-6357 &      0.258 $\pm$ 0.001 &      0.014 $\pm$ 0.001 &      0.259 $\pm$ 0.001 &          3.0 $\pm$ 0.3 &              Y &     &      (i)abe;(ii)bd;0 \\
     J1410-6132 &        0.15 $\pm$ 0.04 &        0.00 $\pm$ 0.04 &        0.18 $\pm$ 0.04 &             2 $\pm$ 16 &              Y &   * &                      \\
     J1412-6145 &        0.58 $\pm$ 0.02 &        0.10 $\pm$ 0.01 &        0.61 $\pm$ 0.02 &             10 $\pm$ 1 &                &   * &                      \\
     J1413-6141 &        0.15 $\pm$ 0.02 &        0.04 $\pm$ 0.02 &        0.22 $\pm$ 0.02 &             16 $\pm$ 9 &                &   * &                      \\
     J1418-3921 &        0.08 $\pm$ 0.01 &        0.03 $\pm$ 0.01 &        0.13 $\pm$ 0.01 &            23 $\pm$ 10 &                &     &                      \\
     J1420-6048 &        0.81 $\pm$ 0.02 &        0.21 $\pm$ 0.02 &        0.89 $\pm$ 0.02 &             15 $\pm$ 1 &                &     &                      \\
     J1424-5822 &      0.181 $\pm$ 0.009 &      0.040 $\pm$ 0.009 &      0.221 $\pm$ 0.009 &             12 $\pm$ 3 &                &     &                      \\
     J1428-5530 &      0.184 $\pm$ 0.001 &      0.015 $\pm$ 0.001 &      0.189 $\pm$ 0.001 &          4.7 $\pm$ 0.4 &              Y &     &      (i)abcf;(ii)b;2 \\
     J1430-6623 &    0.1475 $\pm$ 0.0008 &    0.0391 $\pm$ 0.0007 &    0.1590 $\pm$ 0.0008 &         14.8 $\pm$ 0.3 &              Y &     &       (i)af;(ii)cd;1 \\
     J1435-5954 &      0.101 $\pm$ 0.009 &      0.014 $\pm$ 0.009 &      0.122 $\pm$ 0.009 &              8 $\pm$ 5 &                &     &                      \\
     J1452-6036 &      0.405 $\pm$ 0.007 &      0.090 $\pm$ 0.007 &      0.429 $\pm$ 0.007 &             13 $\pm$ 1 &                &   * &                      \\
     J1453-6413 &    0.2933 $\pm$ 0.0006 &    0.0675 $\pm$ 0.0006 &    0.3083 $\pm$ 0.0006 &         13.0 $\pm$ 0.1 &              Y &     &    (i)abdf;(ii)acd;2 \\
     J1456-6843 &    0.1238 $\pm$ 0.0002 &    0.0617 $\pm$ 0.0002 &    0.1470 $\pm$ 0.0003 &         26.5 $\pm$ 0.2 &              Y &     &     (i)abcf;(ii)cd;2 \\
     J1509-5850 &        0.30 $\pm$ 0.04 &        0.07 $\pm$ 0.04 &        0.39 $\pm$ 0.04 &             13 $\pm$ 8 &                &     &                      \\
     J1512-5759 &      0.107 $\pm$ 0.002 &      0.030 $\pm$ 0.002 &      0.121 $\pm$ 0.002 &             16 $\pm$ 1 &              Y &   * &                      \\
     J1513-5908 &        0.92 $\pm$ 0.02 &        0.14 $\pm$ 0.02 &        0.96 $\pm$ 0.02 &              8 $\pm$ 1 &                &     &                      \\
     J1515-5720 &        0.12 $\pm$ 0.03 &        0.07 $\pm$ 0.03 &        0.21 $\pm$ 0.03 &            30 $\pm$ 20 &                &     &                      \\
     J1522-5829 &      0.295 $\pm$ 0.002 &      0.056 $\pm$ 0.002 &      0.311 $\pm$ 0.002 &         10.7 $\pm$ 0.4 &              Y &   * &                      \\
     J1524-5625 &        0.67 $\pm$ 0.02 &        0.05 $\pm$ 0.01 &        0.70 $\pm$ 0.02 &              4 $\pm$ 1 &                &     &                      \\
     J1524-5706 &        0.49 $\pm$ 0.02 &        0.08 $\pm$ 0.02 &        0.53 $\pm$ 0.02 &              9 $\pm$ 2 &                &     &                      \\
     J1530-5327 &      0.151 $\pm$ 0.009 &      0.020 $\pm$ 0.009 &        0.17 $\pm$ 0.01 &              7 $\pm$ 4 &                &     &                      \\
     J1531-5610 &        0.83 $\pm$ 0.02 &        0.07 $\pm$ 0.01 &        0.85 $\pm$ 0.02 &              5 $\pm$ 1 &                &     &                      \\
     J1534-5334 &      0.091 $\pm$ 0.002 &      0.061 $\pm$ 0.002 &      0.128 $\pm$ 0.002 &             34 $\pm$ 2 &              Y &     &       (i)ab;(ii)cd;1 \\
     J1534-5405 &      0.178 $\pm$ 0.007 &      0.038 $\pm$ 0.007 &      0.206 $\pm$ 0.007 &             12 $\pm$ 3 &                &     &                      \\
     J1535-4114 &      0.494 $\pm$ 0.005 &      0.026 $\pm$ 0.004 &      0.501 $\pm$ 0.005 &          3.0 $\pm$ 0.5 &              Y &     &       (i)acdf;(ii);1 \\
     J1536-5433 &      0.177 $\pm$ 0.006 &      0.029 $\pm$ 0.006 &      0.197 $\pm$ 0.006 &              9 $\pm$ 2 &                &     &                      \\
     J1539-5626 &      0.336 $\pm$ 0.003 &      0.059 $\pm$ 0.003 &      0.355 $\pm$ 0.003 &          9.9 $\pm$ 0.5 &              Y &   * &                      \\
     J1541-5535 &        0.49 $\pm$ 0.03 &        0.07 $\pm$ 0.03 &        0.54 $\pm$ 0.03 &              8 $\pm$ 3 &                &     &                      \\
     J1543-5459 &        0.21 $\pm$ 0.01 &        0.21 $\pm$ 0.01 &        0.39 $\pm$ 0.02 &            45 $\pm$ 11 &                &   * &                      \\
     J1544-5308 &      0.171 $\pm$ 0.002 &      0.071 $\pm$ 0.002 &      0.209 $\pm$ 0.002 &             22 $\pm$ 1 &              Y &     &       (i)adf;(ii)d;2 \\
     J1548-5607 &        0.73 $\pm$ 0.01 &        0.03 $\pm$ 0.01 &        0.73 $\pm$ 0.01 &          2.3 $\pm$ 0.8 &                &   * &                      \\
     J1549-4848 &        0.17 $\pm$ 0.01 &        0.03 $\pm$ 0.01 &        0.18 $\pm$ 0.01 &              9 $\pm$ 3 &                &     &                      \\
 J1549-4848 (i) &        0.30 $\pm$ 0.02 &        0.02 $\pm$ 0.02 &        0.34 $\pm$ 0.02 &              5 $\pm$ 4 &                &     &                      \\
     J1555-3134 &      0.140 $\pm$ 0.003 &      0.048 $\pm$ 0.003 &      0.163 $\pm$ 0.003 &             19 $\pm$ 1 &              Y &     &     (i)abdf;(ii)bd;2 \\
     J1557-4258 &      0.297 $\pm$ 0.004 &      0.131 $\pm$ 0.004 &      0.352 $\pm$ 0.004 &         23.8 $\pm$ 0.9 &                &     &                      \\
     J1559-4438 &    0.3705 $\pm$ 0.0004 &    0.0999 $\pm$ 0.0004 &    0.3974 $\pm$ 0.0004 &       15.08 $\pm$ 0.07 &              Y &     &    (i)abcf;(ii)acd;2 \\
     J1600-5044 &    0.1843 $\pm$ 0.0007 &    0.2616 $\pm$ 0.0007 &    0.3349 $\pm$ 0.0007 &             55 $\pm$ 1 &              Y &   * &                      \\
     J1600-5751 &      0.148 $\pm$ 0.007 &      0.041 $\pm$ 0.007 &      0.184 $\pm$ 0.007 &             15 $\pm$ 3 &                &     &                      \\
     J1602-5100 &      0.195 $\pm$ 0.002 &      0.079 $\pm$ 0.002 &      0.230 $\pm$ 0.002 &         22.0 $\pm$ 0.6 &              Y &     &    (i)abc;(ii)abcd;2 \\
     J1603-5657 &      0.399 $\pm$ 0.009 &      0.181 $\pm$ 0.008 &      0.470 $\pm$ 0.009 &             24 $\pm$ 2 &                &     &                      \\
     J1604-4909 &      0.114 $\pm$ 0.002 &      0.033 $\pm$ 0.002 &      0.126 $\pm$ 0.002 &             16 $\pm$ 1 &              Y &     &      (i)abf;(ii)cd;2 \\
     J1605-5257 &      0.361 $\pm$ 0.001 &      0.041 $\pm$ 0.001 &      0.366 $\pm$ 0.001 &          6.5 $\pm$ 0.2 &              Y &     &      (i)abce;(ii)a;2 \\
     J1611-5209 &      0.182 $\pm$ 0.007 &      0.032 $\pm$ 0.007 &      0.198 $\pm$ 0.007 &             10 $\pm$ 2 &              Y &     &      (i)abe;(ii)cd;1 \\
 J1611-5209 (i) &        0.19 $\pm$ 0.07 &        0.05 $\pm$ 0.07 &        0.30 $\pm$ 0.07 &            16 $\pm$ 23 &              Y &     &      (i)abe;(ii)cd;1 \\
     J1613-4714 &      0.145 $\pm$ 0.006 &      0.031 $\pm$ 0.006 &      0.160 $\pm$ 0.006 &             12 $\pm$ 2 &                &     &                      \\
     J1614-5048 &      0.725 $\pm$ 0.007 &      0.192 $\pm$ 0.006 &      0.773 $\pm$ 0.007 &         14.8 $\pm$ 0.5 &                &   * &                      \\
     J1623-4256 &      0.258 $\pm$ 0.005 &      0.047 $\pm$ 0.005 &      0.277 $\pm$ 0.005 &             10 $\pm$ 1 &                &     &                      \\
     J1626-4537 &      0.061 $\pm$ 0.009 &      0.025 $\pm$ 0.009 &      0.100 $\pm$ 0.009 &            23 $\pm$ 11 &                &     &                      \\
     J1630-4733 &      0.302 $\pm$ 0.005 &      0.157 $\pm$ 0.005 &      0.384 $\pm$ 0.006 &             27 $\pm$ 1 &                &   * &                      \\
     J1632-4621 &        0.27 $\pm$ 0.01 &        0.04 $\pm$ 0.01 &        0.28 $\pm$ 0.01 &              8 $\pm$ 3 &                &   * &                      \\
     J1633-4453 &      0.130 $\pm$ 0.005 &      0.016 $\pm$ 0.005 &      0.142 $\pm$ 0.005 &              7 $\pm$ 2 &                &   * &                      \\
     J1633-5015 &      0.293 $\pm$ 0.002 &      0.121 $\pm$ 0.002 &      0.328 $\pm$ 0.002 &         22.5 $\pm$ 0.4 &              Y &   * &                      \\
 J1637-4553 (i) &        0.24 $\pm$ 0.07 &        0.12 $\pm$ 0.07 &        0.40 $\pm$ 0.08 &            27 $\pm$ 24 &                &     &                      \\
     J1637-4553 &        0.88 $\pm$ 0.01 &      0.021 $\pm$ 0.008 &        0.89 $\pm$ 0.01 &          1.4 $\pm$ 0.6 &                &     &                      \\
     J1637-4642 &        0.89 $\pm$ 0.03 &        0.08 $\pm$ 0.02 &        0.93 $\pm$ 0.03 &              5 $\pm$ 1 &                &     &                      \\
     J1638-4417 &        0.75 $\pm$ 0.04 &        0.03 $\pm$ 0.03 &        0.77 $\pm$ 0.04 &              3 $\pm$ 2 &                &     &                      \\
     J1638-4608 &        0.65 $\pm$ 0.02 &        0.05 $\pm$ 0.02 &        0.67 $\pm$ 0.02 &              4 $\pm$ 2 &                &     &                      \\
     J1638-4725 &        0.08 $\pm$ 0.02 &        0.03 $\pm$ 0.02 &        0.15 $\pm$ 0.02 &            23 $\pm$ 18 &                &   * &                      \\
     J1640-4715 &        0.18 $\pm$ 0.01 &        0.02 $\pm$ 0.01 &        0.21 $\pm$ 0.01 &              7 $\pm$ 4 &                &   * &                      \\
     J1643-4505 &        0.58 $\pm$ 0.03 &        0.34 $\pm$ 0.02 &        0.76 $\pm$ 0.03 &             30 $\pm$ 4 &                &     &                      \\
     J1644-4559 &  0.21939 $\pm$ 0.00008 &  0.03564 $\pm$ 0.00008 &  0.22439 $\pm$ 0.00008 &        9.23 $\pm$ 0.02 &              Y &   * &                      \\
     J1645-0317 &    0.1324 $\pm$ 0.0005 &    0.0324 $\pm$ 0.0005 &    0.1422 $\pm$ 0.0005 &         13.8 $\pm$ 0.2 &              Y &     &    (i)abc;(ii)abcd;1 \\
     J1646-4346 &        0.43 $\pm$ 0.01 &        0.02 $\pm$ 0.01 &        0.45 $\pm$ 0.01 &              3 $\pm$ 2 &                &   * &                      \\
     J1646-6831 &      0.383 $\pm$ 0.003 &      0.079 $\pm$ 0.003 &      0.410 $\pm$ 0.003 &         11.6 $\pm$ 0.5 &                &     &                      \\
     J1648-3256 &      0.115 $\pm$ 0.006 &      0.068 $\pm$ 0.006 &      0.167 $\pm$ 0.007 &             30 $\pm$ 5 &                &     &                      \\
     J1648-4611 &        0.70 $\pm$ 0.02 &        0.15 $\pm$ 0.01 &        0.75 $\pm$ 0.02 &             12 $\pm$ 1 &                &     &                      \\
     J1649-3805 &        0.16 $\pm$ 0.01 &        0.06 $\pm$ 0.01 &        0.21 $\pm$ 0.01 &             20 $\pm$ 5 &                &     &                      \\
     J1649-4653 &        0.52 $\pm$ 0.03 &        0.05 $\pm$ 0.03 &        0.56 $\pm$ 0.03 &              6 $\pm$ 3 &                &     &                      \\
     J1650-4502 &        0.73 $\pm$ 0.01 &        0.05 $\pm$ 0.01 &        0.75 $\pm$ 0.01 &          4.3 $\pm$ 0.8 &                &   * &                      \\
     J1650-4921 &        0.75 $\pm$ 0.02 &        0.26 $\pm$ 0.01 &        0.81 $\pm$ 0.02 &             19 $\pm$ 1 &                &     &                      \\
     J1651-4246 &      0.444 $\pm$ 0.001 &      0.127 $\pm$ 0.001 &      0.497 $\pm$ 0.001 &         16.0 $\pm$ 0.2 &              Y &   * &                      \\
     J1651-5222 &      0.174 $\pm$ 0.002 &      0.075 $\pm$ 0.002 &      0.209 $\pm$ 0.002 &             23 $\pm$ 1 &              Y &     &      (i)ab;(ii)bcd;2 \\
     J1651-5255 &      0.368 $\pm$ 0.004 &      0.079 $\pm$ 0.004 &      0.394 $\pm$ 0.004 &         12.1 $\pm$ 0.7 &                &     &                      \\
     J1652-2404 &      0.112 $\pm$ 0.006 &      0.053 $\pm$ 0.006 &      0.142 $\pm$ 0.006 &             25 $\pm$ 4 &                &     &                      \\
     J1653-3838 &      0.360 $\pm$ 0.003 &      0.054 $\pm$ 0.003 &      0.370 $\pm$ 0.003 &          8.5 $\pm$ 0.5 &              Y &     &         (i);(ii)ad;1 \\
     J1653-4249 &      0.161 $\pm$ 0.007 &      0.022 $\pm$ 0.007 &      0.176 $\pm$ 0.007 &              8 $\pm$ 3 &                &   * &                      \\
     J1700-3312 &      0.361 $\pm$ 0.009 &      0.120 $\pm$ 0.008 &      0.407 $\pm$ 0.009 &             18 $\pm$ 2 &                &     &                      \\
     J1701-3726 &      0.313 $\pm$ 0.003 &      0.187 $\pm$ 0.003 &      0.398 $\pm$ 0.003 &         30.9 $\pm$ 0.9 &              Y &     &   (i)abcde;(ii)bcd;2 \\
     J1701-4533 &      0.268 $\pm$ 0.005 &      0.104 $\pm$ 0.005 &      0.306 $\pm$ 0.005 &             21 $\pm$ 1 &                &   * &                      \\
     J1702-4128 &        0.60 $\pm$ 0.02 &        0.04 $\pm$ 0.01 &        0.62 $\pm$ 0.02 &              4 $\pm$ 1 &                &     &                      \\
     J1702-4310 &        0.96 $\pm$ 0.02 &        0.11 $\pm$ 0.01 &        0.99 $\pm$ 0.02 &          6.6 $\pm$ 0.8 &                &     &                      \\
     J1703-3241 &      0.429 $\pm$ 0.001 &      0.089 $\pm$ 0.001 &      0.443 $\pm$ 0.001 &         11.7 $\pm$ 0.2 &              Y &     &     (i)acde;(ii)cd;2 \\
     J1703-4851 &      0.135 $\pm$ 0.008 &      0.062 $\pm$ 0.008 &      0.183 $\pm$ 0.008 &             25 $\pm$ 4 &                &     &                      \\
     J1705-1906 &      0.473 $\pm$ 0.003 &      0.186 $\pm$ 0.002 &      0.530 $\pm$ 0.003 &         21.5 $\pm$ 0.4 &              Y &     &        (i)ae;(ii)d;2 \\
 J1705-1906 (i) &        0.81 $\pm$ 0.01 &      0.180 $\pm$ 0.009 &        0.84 $\pm$ 0.01 &         12.6 $\pm$ 0.7 &              Y &     &        (i)ae;(ii)d;2 \\
     J1705-3423 &      0.130 $\pm$ 0.002 &      0.031 $\pm$ 0.002 &      0.146 $\pm$ 0.003 &             13 $\pm$ 1 &              Y &     &        (i)f;(ii)bd;0 \\
     J1705-3950 &        0.80 $\pm$ 0.01 &      0.266 $\pm$ 0.009 &        0.90 $\pm$ 0.01 &         18.3 $\pm$ 0.8 &                &     &                      \\
     J1707-4053 &      0.240 $\pm$ 0.003 &      0.055 $\pm$ 0.003 &      0.257 $\pm$ 0.003 &         13.0 $\pm$ 0.7 &                &   * &                      \\
     J1707-4729 &      0.309 $\pm$ 0.007 &      0.083 $\pm$ 0.007 &      0.349 $\pm$ 0.007 &             15 $\pm$ 1 &                &   * &                      \\
     J1708-3426 &      0.166 $\pm$ 0.005 &      0.088 $\pm$ 0.005 &      0.217 $\pm$ 0.005 &             28 $\pm$ 2 &                &     &                      \\
     J1709-1640 &      0.104 $\pm$ 0.001 &      0.014 $\pm$ 0.001 &      0.110 $\pm$ 0.001 &          7.7 $\pm$ 0.7 &              Y &     &       (i)abe;(ii)d;0 \\
     J1709-4429 &      0.950 $\pm$ 0.002 &      0.213 $\pm$ 0.002 &      0.980 $\pm$ 0.002 &         12.7 $\pm$ 0.1 &              Y &     &           (i);(ii);0 \\
     J1715-3903 &        0.89 $\pm$ 0.03 &        0.09 $\pm$ 0.02 &        0.92 $\pm$ 0.03 &              6 $\pm$ 1 &                &     &                      \\
     J1715-4034 &      0.224 $\pm$ 0.007 &      0.033 $\pm$ 0.007 &      0.243 $\pm$ 0.007 &              8 $\pm$ 2 &                &     &                      \\
     J1716-4005 &        0.08 $\pm$ 0.01 &        0.03 $\pm$ 0.01 &        0.12 $\pm$ 0.01 &             19 $\pm$ 9 &                &   * &                      \\
     J1717-3425 &      0.092 $\pm$ 0.003 &      0.071 $\pm$ 0.003 &      0.140 $\pm$ 0.003 &             38 $\pm$ 4 &                &   * &                      \\
     J1718-3825 &        0.95 $\pm$ 0.02 &        0.14 $\pm$ 0.01 &        0.99 $\pm$ 0.02 &          8.2 $\pm$ 0.7 &                &     &                      \\
     J1719-4006 &      0.080 $\pm$ 0.009 &      0.042 $\pm$ 0.009 &      0.121 $\pm$ 0.009 &             28 $\pm$ 9 &                &     &                      \\
     J1720-2933 &      0.152 $\pm$ 0.007 &      0.050 $\pm$ 0.007 &      0.185 $\pm$ 0.007 &             18 $\pm$ 3 &                &     &                      \\
     J1721-3532 &      0.081 $\pm$ 0.003 &      0.051 $\pm$ 0.003 &      0.130 $\pm$ 0.003 &             32 $\pm$ 4 &                &   * &                      \\
     J1722-3207 &      0.233 $\pm$ 0.002 &      0.031 $\pm$ 0.002 &      0.241 $\pm$ 0.002 &          7.6 $\pm$ 0.5 &              Y &     &    (i)abcef;(ii)cd;1 \\
     J1722-3632 &      0.193 $\pm$ 0.007 &      0.041 $\pm$ 0.007 &      0.218 $\pm$ 0.007 &             12 $\pm$ 2 &                &     &                      \\
 J1722-3712 (i) &        0.86 $\pm$ 0.02 &        0.03 $\pm$ 0.02 &        0.87 $\pm$ 0.02 &              2 $\pm$ 1 &              Y &     &           (i);(ii);0 \\
     J1722-3712 &      0.377 $\pm$ 0.002 &      0.122 $\pm$ 0.002 &      0.403 $\pm$ 0.002 &         18.0 $\pm$ 0.4 &              Y &     &           (i);(ii);0 \\
     J1723-3659 &      0.544 $\pm$ 0.008 &      0.175 $\pm$ 0.007 &      0.597 $\pm$ 0.008 &         17.8 $\pm$ 0.9 &                &     &                      \\
     J1727-2739 &      0.489 $\pm$ 0.004 &      0.045 $\pm$ 0.004 &      0.496 $\pm$ 0.004 &          5.3 $\pm$ 0.5 &                &     &                      \\
     J1730-3350 &      0.572 $\pm$ 0.007 &      0.023 $\pm$ 0.006 &      0.582 $\pm$ 0.007 &          2.3 $\pm$ 0.6 &                &   * &                      \\
     J1731-4744 &      0.170 $\pm$ 0.002 &      0.018 $\pm$ 0.002 &      0.173 $\pm$ 0.002 &          6.0 $\pm$ 0.8 &              Y &     &     (i)abce;(ii)cd;2 \\
     J1733-2228 &      0.195 $\pm$ 0.004 &      0.090 $\pm$ 0.004 &      0.246 $\pm$ 0.004 &             25 $\pm$ 2 &                &     &                      \\
     J1733-3716 &      0.791 $\pm$ 0.007 &      0.129 $\pm$ 0.006 &      0.832 $\pm$ 0.007 &          9.3 $\pm$ 0.4 &              Y &     &          (i);(ii)a;1 \\
     J1735-0724 &      0.260 $\pm$ 0.005 &      0.041 $\pm$ 0.005 &      0.277 $\pm$ 0.005 &              9 $\pm$ 1 &                &     &                      \\
     J1737-3137 &        0.62 $\pm$ 0.02 &        0.21 $\pm$ 0.02 &        0.71 $\pm$ 0.02 &             19 $\pm$ 2 &                &     &                      \\
     J1738-3211 &      0.177 $\pm$ 0.005 &      0.049 $\pm$ 0.005 &      0.193 $\pm$ 0.005 &             15 $\pm$ 2 &                &     &                      \\
 J1739-2903 (i) &        0.36 $\pm$ 0.01 &      0.044 $\pm$ 0.009 &        0.37 $\pm$ 0.01 &              7 $\pm$ 2 &              Y &     &      (i)a;(ii)abcd;2 \\
     J1739-2903 &      0.086 $\pm$ 0.004 &      0.031 $\pm$ 0.004 &      0.103 $\pm$ 0.004 &             20 $\pm$ 3 &              Y &     &      (i)a;(ii)abcd;2 \\
     J1739-3023 &        0.80 $\pm$ 0.01 &        0.04 $\pm$ 0.01 &        0.81 $\pm$ 0.01 &          2.8 $\pm$ 0.7 &                &     &                      \\
     J1739-3131 &      0.123 $\pm$ 0.004 &      0.022 $\pm$ 0.004 &      0.140 $\pm$ 0.004 &             10 $\pm$ 2 &                &   * &                      \\
     J1740-3015 &      0.834 $\pm$ 0.003 &      0.364 $\pm$ 0.002 &      0.945 $\pm$ 0.003 &         23.6 $\pm$ 0.2 &              Y &     &        (i)aef;(ii);1 \\
     J1741-2733 &      0.174 $\pm$ 0.008 &      0.032 $\pm$ 0.008 &      0.199 $\pm$ 0.008 &             10 $\pm$ 3 &                &     &                      \\
     J1741-3016 &        0.37 $\pm$ 0.01 &        0.03 $\pm$ 0.01 &        0.39 $\pm$ 0.01 &              5 $\pm$ 2 &                &     &                      \\
     J1741-3927 &      0.237 $\pm$ 0.003 &      0.034 $\pm$ 0.003 &      0.246 $\pm$ 0.003 &          8.3 $\pm$ 0.6 &              Y &     &    (i)abef;(ii)abd;2 \\
     J1743-3150 &        0.20 $\pm$ 0.01 &        0.01 $\pm$ 0.01 &        0.20 $\pm$ 0.01 &              2 $\pm$ 3 &                &     &                      \\
     J1745-3040 &      0.489 $\pm$ 0.001 &      0.040 $\pm$ 0.001 &      0.493 $\pm$ 0.001 &          4.7 $\pm$ 0.1 &              Y &     &      (i)ae;(ii)acd;2 \\
     J1749-3002 &      0.307 $\pm$ 0.006 &      0.113 $\pm$ 0.005 &      0.372 $\pm$ 0.006 &             20 $\pm$ 1 &                &     &                      \\
     J1750-3157 &        0.28 $\pm$ 0.02 &        0.07 $\pm$ 0.01 &        0.34 $\pm$ 0.02 &             14 $\pm$ 3 &                &     &                      \\
     J1751-3323 &        0.30 $\pm$ 0.01 &        0.04 $\pm$ 0.01 &        0.32 $\pm$ 0.01 &              7 $\pm$ 2 &                &     &                      \\
     J1751-4657 &      0.159 $\pm$ 0.001 &      0.088 $\pm$ 0.001 &      0.199 $\pm$ 0.001 &         28.9 $\pm$ 0.6 &              Y &     &     (i)bcde;(ii)cd;1 \\
     J1752-2806 &    0.0951 $\pm$ 0.0004 &    0.0716 $\pm$ 0.0004 &    0.1266 $\pm$ 0.0004 &         37.0 $\pm$ 0.5 &              Y &     &    (i)abce;(ii)bcd;2 \\
     J1757-2421 &      0.246 $\pm$ 0.002 &      0.035 $\pm$ 0.002 &      0.254 $\pm$ 0.002 &          8.1 $\pm$ 0.6 &              Y &     &      (i)abd;(ii)ad;1 \\
     J1801-2304 &        0.05 $\pm$ 0.01 &        0.01 $\pm$ 0.01 &        0.07 $\pm$ 0.01 &            17 $\pm$ 16 &              Y &   * &                      \\
     J1801-2451 &        0.80 $\pm$ 0.01 &      0.181 $\pm$ 0.009 &        0.84 $\pm$ 0.01 &         12.7 $\pm$ 0.7 &                &     &                      \\
     J1803-2137 &      0.577 $\pm$ 0.004 &      0.258 $\pm$ 0.003 &      0.662 $\pm$ 0.004 &         24.1 $\pm$ 0.5 &              Y &     &          (i)a;(ii);1 \\
     J1807-0847 &    0.2487 $\pm$ 0.0007 &    0.0886 $\pm$ 0.0007 &    0.2752 $\pm$ 0.0007 &         19.6 $\pm$ 0.2 &              Y &     &    (i)abf;(ii)abcd;2 \\
     J1809-1917 &        0.85 $\pm$ 0.01 &        0.27 $\pm$ 0.01 &        0.94 $\pm$ 0.01 &         17.6 $\pm$ 0.9 &                &     &                      \\
     J1816-2650 &      0.310 $\pm$ 0.006 &      0.085 $\pm$ 0.006 &      0.343 $\pm$ 0.006 &             15 $\pm$ 1 &                &     &                      \\
     J1817-3618 &      0.171 $\pm$ 0.005 &      0.120 $\pm$ 0.005 &      0.244 $\pm$ 0.005 &             35 $\pm$ 3 &                &     &                      \\
     J1817-3837 &      0.139 $\pm$ 0.006 &      0.012 $\pm$ 0.006 &      0.155 $\pm$ 0.006 &              5 $\pm$ 2 &              Y &     &         (i)f;(ii)d;2 \\
     J1820-0427 &      0.240 $\pm$ 0.002 &      0.095 $\pm$ 0.002 &      0.272 $\pm$ 0.002 &         21.5 $\pm$ 0.5 &              Y &     &    (i)abef;(ii)bcd;2 \\
     J1822-2256 &      0.210 $\pm$ 0.005 &      0.040 $\pm$ 0.005 &      0.225 $\pm$ 0.005 &             11 $\pm$ 2 &                &     &                      \\
     J1822-4209 &        0.18 $\pm$ 0.02 &        0.02 $\pm$ 0.02 &        0.22 $\pm$ 0.02 &              7 $\pm$ 5 &                &     &                      \\
     J1823-3106 &      0.443 $\pm$ 0.002 &      0.049 $\pm$ 0.002 &      0.452 $\pm$ 0.002 &          6.4 $\pm$ 0.3 &              Y &     &        (i)af;(ii)d;1 \\
     J1824-1945 &      0.210 $\pm$ 0.001 &      0.048 $\pm$ 0.001 &      0.223 $\pm$ 0.001 &         12.9 $\pm$ 0.3 &              Y &     &     (i)ab;(ii)abcd;1 \\
     J1825-0935 &      0.279 $\pm$ 0.002 &      0.036 $\pm$ 0.002 &      0.286 $\pm$ 0.002 &          7.3 $\pm$ 0.4 &              Y &     &          (i)b;(ii);1 \\
 J1825-0935 (i) &        0.10 $\pm$ 0.02 &        0.02 $\pm$ 0.02 &        0.12 $\pm$ 0.02 &             9 $\pm$ 11 &              Y &     &          (i)b;(ii);1 \\
     J1825-1446 &      0.633 $\pm$ 0.006 &      0.020 $\pm$ 0.005 &      0.638 $\pm$ 0.006 &          1.9 $\pm$ 0.4 &                &   * &                      \\
     J1826-1334 &      0.863 $\pm$ 0.008 &      0.308 $\pm$ 0.006 &      0.956 $\pm$ 0.008 &         19.7 $\pm$ 0.5 &                &     &                      \\
 J1828-1101 (i) &        0.27 $\pm$ 0.05 &        0.06 $\pm$ 0.05 &        0.35 $\pm$ 0.05 &            13 $\pm$ 12 &                &   * &                      \\
     J1828-1101 &        0.71 $\pm$ 0.01 &        0.04 $\pm$ 0.01 &        0.73 $\pm$ 0.01 &          2.8 $\pm$ 0.9 &                &   * &                      \\
     J1829-1751 &      0.342 $\pm$ 0.001 &      0.090 $\pm$ 0.001 &      0.380 $\pm$ 0.001 &         14.8 $\pm$ 0.2 &              Y &     &     (i)abd;(ii)acd;2 \\
     J1830-1059 &      0.857 $\pm$ 0.007 &      0.167 $\pm$ 0.005 &      0.880 $\pm$ 0.007 &         11.0 $\pm$ 0.4 &              Y &     &          (i)b;(ii);1 \\
     J1832-0827 &      0.218 $\pm$ 0.003 &      0.028 $\pm$ 0.003 &      0.229 $\pm$ 0.003 &          7.3 $\pm$ 0.9 &              Y &     &      (i)ade;(ii)ad;1 \\
     J1833-0827 &      0.205 $\pm$ 0.003 &      0.036 $\pm$ 0.003 &      0.221 $\pm$ 0.003 &             10 $\pm$ 1 &                &   * &                      \\
     J1834-0426 &      0.268 $\pm$ 0.002 &      0.037 $\pm$ 0.002 &      0.281 $\pm$ 0.002 &          7.8 $\pm$ 0.4 &              Y &     &     (i)abf;(ii)acd;2 \\
     J1835-0643 &        0.10 $\pm$ 0.01 &        0.04 $\pm$ 0.01 &        0.16 $\pm$ 0.01 &             20 $\pm$ 9 &                &   * &                      \\
     J1835-1106 &      0.619 $\pm$ 0.006 &      0.066 $\pm$ 0.005 &      0.629 $\pm$ 0.006 &          6.1 $\pm$ 0.5 &              Y &     &          (i)c;(ii);0 \\
     J1841-0345 &        0.94 $\pm$ 0.01 &      0.033 $\pm$ 0.009 &        0.95 $\pm$ 0.01 &          2.0 $\pm$ 0.6 &                &     &                      \\
     J1841-0425 &      0.616 $\pm$ 0.005 &      0.035 $\pm$ 0.004 &      0.624 $\pm$ 0.005 &          3.2 $\pm$ 0.4 &              Y &   * &                      \\
     J1842-0905 &        0.21 $\pm$ 0.01 &        0.06 $\pm$ 0.01 &        0.26 $\pm$ 0.01 &             17 $\pm$ 3 &                &     &                      \\
     J1844-0538 &      0.492 $\pm$ 0.005 &      0.068 $\pm$ 0.004 &      0.506 $\pm$ 0.005 &          7.8 $\pm$ 0.5 &              Y &   * &                      \\
     J1845-0434 &      0.192 $\pm$ 0.003 &      0.066 $\pm$ 0.003 &      0.226 $\pm$ 0.003 &             19 $\pm$ 1 &                &     &                      \\
     J1845-0743 &      0.326 $\pm$ 0.004 &      0.028 $\pm$ 0.004 &      0.337 $\pm$ 0.004 &          5.0 $\pm$ 0.7 &              Y &     &        (i)b;(ii)ac;1 \\
     J1847-0402 &      0.194 $\pm$ 0.003 &      0.049 $\pm$ 0.003 &      0.210 $\pm$ 0.003 &         14.3 $\pm$ 0.9 &              Y &     &     (i)abce;(ii)cd;2 \\
     J1848-0123 &      0.170 $\pm$ 0.001 &      0.020 $\pm$ 0.001 &      0.179 $\pm$ 0.001 &          6.6 $\pm$ 0.5 &              Y &     &    (i)abce;(ii)acd;2 \\
     J1852-0635 &      0.618 $\pm$ 0.003 &      0.011 $\pm$ 0.002 &      0.621 $\pm$ 0.003 &          1.0 $\pm$ 0.2 &              Y &     &         (i)a;(ii)a;1 \\
     J1852-2610 &        0.36 $\pm$ 0.01 &        0.03 $\pm$ 0.01 &        0.40 $\pm$ 0.01 &              4 $\pm$ 2 &                &     &                      \\
     J1853-0004 &        0.86 $\pm$ 0.03 &        0.04 $\pm$ 0.02 &        0.88 $\pm$ 0.03 &              3 $\pm$ 1 &                &     &                      \\
     J1900-2600 &      0.340 $\pm$ 0.001 &      0.147 $\pm$ 0.001 &      0.402 $\pm$ 0.001 &         23.4 $\pm$ 0.2 &              Y &     &   (i)abcdf;(ii)bcd;2 \\
     J1913-0440 &      0.136 $\pm$ 0.002 &      0.040 $\pm$ 0.002 &      0.147 $\pm$ 0.002 &         16.4 $\pm$ 0.8 &              Y &     &    (i)abce;(ii)abd;2 \\
     J1941-2602 &      0.514 $\pm$ 0.004 &      0.113 $\pm$ 0.004 &      0.544 $\pm$ 0.004 &         12.4 $\pm$ 0.5 &              Y &     &          (i);(ii)c;2 \\
     J2006-0807 &      0.433 $\pm$ 0.009 &      0.062 $\pm$ 0.008 &      0.461 $\pm$ 0.009 &              8 $\pm$ 1 &                &     &                      \\
     J2048-1616 &    0.3900 $\pm$ 0.0007 &    0.0467 $\pm$ 0.0006 &    0.3949 $\pm$ 0.0007 &        6.83 $\pm$ 0.09 &              Y &     &       (i)adf;(ii)d;1 \\
     J2330-2005 &      0.147 $\pm$ 0.003 &      0.070 $\pm$ 0.003 &      0.188 $\pm$ 0.003 &             25 $\pm$ 1 &              Y &     &       (i)abc;(ii)c;2 \\
     J2346-0609 &      0.451 $\pm$ 0.003 &      0.102 $\pm$ 0.003 &      0.468 $\pm$ 0.003 &         12.8 $\pm$ 0.3 &              Y &     &         (i)c;(ii)a;1 \\
\end{longtable}

\twocolumn

% Don't change these lines
\bsp	% typesetting comment
\label{lastpage}
\end{document}